\documentclass[aps,twocolumn]{revtex4-1}
\usepackage{amsmath,amssymb}
\usepackage{slashed}
\usepackage{graphicx}
\usepackage{bm}
\usepackage[T1]{fontenc}
\usepackage[utf8]{inputenc}
\usepackage{gauss} 
\usepackage[top=1.0in,bottom=1.0in,left=1.0in,right=1.0in]{geometry}
\usepackage[colorlinks,linkcolor=blue,citecolor=blue]{hyperref}
\usepackage{comment}
\newcommand{\be}{\begin{equation}}
\newcommand{\ee}{\end{equation}}
\newcommand{\bea}{\begin{eqnarray}}
\newcommand{\eea}{\end{eqnarray}}
\newcommand{\ba}{\begin{eqnarray}}
\newcommand{\ea}{\end{eqnarray}}

\begin{document}

\title{
Wave functions of multiquark hadrons \\ from representations of the symmetry groups $S_n$
}

\author{Nicholas Miesch}
\email{nicholas.miesch@stonybrook.edu}
\affiliation{Center for Nuclear Theory, Department of Physics and Astronomy, Stony Brook University, Stony Brook, New York 11794--3800, USA}

\author{Edward Shuryak}
\email{edward.shuryak@stonybrook.edu}
\affiliation{Center for Nuclear Theory, Department of Physics and Astronomy, Stony Brook University, Stony Brook, New York 11794--3800, USA}


\begin{abstract} 
Construction of the wave functions of multiquark hadrons by traditional method based on tensor products of colors, flavors, spins (and orbital) parts becomes quite complex
when quark numbers grow $n=5,6...12$, as it gets difficult to
satisfy  requirements of Fermi statistics. Our novel approach
is focused directly on representations of the permutation symmetry generators. After showing how
$C_3$ is manifested in the wave functions of (excited)  baryons, we use it to construct the wave functions for a set of pentaquarks and  hexaquarks (n=5,6). We also have some partial results for larger systems, with $n=9$ and 12, and even beyond that as far as $n=24$.  
 \end{abstract}

\maketitle

\section{Introduction}
\subsection{Outline}
The main issue discussed in this work is a very old one: how to
construct hadronic wave functions (WFs) which are totally antisymmetric under
quark permutations, as Fermi statistics requires. While for mesons and some baryons it is 
rather simple task, in general it is not so since 
there arise parts of the WFs which
 possess   ``mixed symmetries",  for  color,flavor,spin (and orbital part,
at nonzero orbital momentum $L$). To construct WFs, by
linear superposition of these parts, with appropriate permutation symmetries, rapidly becomes difficult, as the number of constituents grows.

 The simplest baryons like $\Delta$ (and other  members of $SU(3)$ decuplet) 
have WF factorized into a product of $antisymmetric$ color and $symmetric$ flavor and spin WFs as  $I,S=3/2$. Its wave function is thus represented by a single $monom$ $(u^\uparrow u^\uparrow u^\uparrow)$.

Yet already the flavor-spin WF of nucleons (and other  members of $SU(3)$ octet) 
get more complicated since their quantum numbers ($I,S=1/2)$ prevent construction of separate permutaion-symmetric WFs for spin and flavor. One can
easily construct spin-1/2 WFs symmetric (or antisymmetric) under interchange
of, say, quarks 1 and 2 ($P_{12}$), yet they are not symmetric under other (e.g. 
$P_{23},P_{13}$ permutations. One has to  look at superposition 
of $S_{sym}F_{sym}$ and $S_{asym}F_{asym}$ terms, symmetric under $P_{12}$, and check which one 
 would be symmetric under other permutations. 

The method we developed in this paper stems from
 our recent paper \cite{Miesch:2023hvl} in which we addressed the WFs of
(negative parity) $L=1$ and $L=2$ nucleon resonances. Their WFs   include orbital
factors, on top of spin and flavor variables. Those depend on angles of the
(modified) Jacobi coordinates and also  have mixed
permutation symmetries,  thus the problem
 gets even more complicated. With more parts of flavor-spin-orbital combinations, we solved the problem and
obtained explicit WFs by first building ``tensor cubed" representations of the generators of the symmetry group $S_3$. We also advocated usage of spin-tensor notations, in space of all existing ``monoms", using Mathematica.

In this paper we  generalize this method further, to multiquark hadrons, building  representations of higher symmetry groups $S_n$, with $n=4,5,6..$ respectively, see
section \ref{sec_procedure}.
We focus mostly on $n=6$, hexaquarks (or dibaryons) and
 pentaquarks. Starting from spin-tensor notations with all monoms,
combining color, flavor,spin etc indices, one finds that their dimensions
rapidly become  quite large -- e.g. for n=6 $u,d$ quarks there are $3^6\cdot 2^6 \cdot 2^6=2985984$ monoms, and the square of that
for the n=12 case. Those become hard to handle even
using Mathematica due to computer memory limitations. 

Yet the actually needed dimensions of pertinent vectors and matrices can be greatly reduced by
using what we call  a ``good basis" of linearly independent and mutually orthogonal combinations,as we will detail below. For example, for hexaquark with maximal spin, the reduction of color-flavor monom space $3^6\cdot 2^6$ get reduced to just $ 5\cdot 5$ space.

Another important idea (well known in mathematics) is that there is no need to consider all $n!$ permutations
 of the symmetry group $C_n$, but build  matrix forms  only for its $two$ basic
generators, namely $P_{12}, P_{cycle}$. After diagonalization of those, it is rather easy to see if they
possess $common$ eigenvectors with the needed symmetry. If they do,
those are the WF one is looking for.

Our approach is much more direct than standard approach,
based on  building subsequent representations  starting from spin, flavor $SU(2)$ and color $SU(3)$ groups via tensor products, eventually reaching the (tensor product) power $n$ (for $n$ quarks) starts with selecting certain pairs, then pairs of pairs etc. The very first step on this road -- randomly selecting the original pairs -- is arbitrary and contrary to
the final goal, keeping certain global permutation intact.

We instead of usual quantum numbers focus on representations
of symmetry groups $S_n$. It also starts with tensor product
of certain number of its representations. 
Completing the historic Introduction, let us add what (we learned) from its mathematical history.

A given Young Tableau is a simultaneous description of both an $SU(N)$ and an $S_n$ representation, the $S_n$ representations of the pieces of the wavefunction for each sector match the $SU(N)$ one, except instead of neglecting the $N$-tall stacks we include them.  So, if $\lambda_c$ is the Young Tableau corresponding to the $SU(3)$ representation for color, $\lambda_f$ is the Young Tableau corresponding to the $SU(2)$ representation for flavor, and so on, then 
their tensor product should be decomposed into a sum of
irreducible representations
\be
    \lambda_c\otimes\lambda_f\otimes\dots=\bigoplus_{\mu\in \text{ reps of }S_n}C^\mu_{\lambda_c\lambda_f\dots}\mu
\ee
with "Clebsch-Gordon" coefficients. (In mathematics these constants are typically called Kronecker Coefficients. )
The question of whether there exists a state with the correct Fermi statistics translates to the question of whether the coefficient $C_{\text{anti} \lambda_c\lambda_f\dots}$ in these  series 
is zero or not. 

  There has been a large amount of mathematical investigation into them, and there exist efficient online resources to find them. We suggest 
  the website \cite{gibsonKronecker}, which can help answering the question of existence of the WFs, even for quark numbers that are too large for our procedure based on Mathematica to handle and find them explicitly (see section \ref{sec_big_permute}).  When these Kronecker coefficient tables return 0 it rules out the possibility of Fermi- statistics-obeying WFs for a given combination of quantum numbers. Needless to say, in all cases we were able
  to do it explicitly, the results agree with the output of this page. For some explanations of how to use \cite{gibsonKronecker}, see appendix \ref{sec_kroneckerTables}.

Completing the general introduction, let us briefly comment on 
selection of particular examples we will discuss. Since we have already discussed baryons and tetraquarks in our previous work \cite{Miesch:2023hvl},
the natural extensions are $penta-$ and, especially, $hexaquarks$ (or 
dibaryons, for general reviews see e.g. \cite{Clement_2017,Bashkanov:2023yca}).
In section \ref{sec_S3_hex} we start showing how all procedures suggested are used in the case of $uuuddd$ hexaquark with maximal spin $S=3$, which is then applied to other spins, flavors and orbital momentum  in section \ref{sec_other_hex}.
Pentaquarks are discussed in section \ref{sec_penta}.
We also discuss challenging applications to larger systems,  with 9 quarks (tribaryons) and 12 quarks (tetrabaryons),   in section \ref{sec_9_12}.

\subsection{Historic remarks} \label{sec_historic}
Hadronic spectroscopy started in 1960's from flavor 
$SU(3)$ symmetry, with Gell-Mann's ``eightfold way" based on its octet
adjoint representation. Then $decuplets$ came, both
algebraically and then experimentally, completed with famous observation of $\Omega^-=sss, I=J=3/2$ hyperon.
Combining flavor with $SU(2)$ spin into the $SU(6)$ group, one was 
able to get ``squared" and ``cubed" representations of it, as needed for basic mesons and baryons. Yet already for baryons with nonzero
orbital momentum $L=1,2..$ it gets so complicated that (e.g.
in classic papers like \cite{Isgur:1978xj}) their WFs enforcing Fermi statistics were not
explicitly constructed. We did so recently in \cite{Miesch:2023hvl}.

This paper deals with technical theoretical issues, aiming at explicit
construction of WFs for
mutiquark hadrons, with $n=4,5,6...12$. One may wonder if
those are actually needed for any real-life applications. 
Indeed, for about 40 years (1965-2005) it was considered to be a purely academic subject, but 
during the last two decades 
discoveries of many multiquark resonances suddenly became numerous, turning 
 the field of hadronic spectroscopy into  a real renaissance. 
 The leader in this direction is the $LHCb$ collaboration,
 making good use of multiple production of $c,b$ quarks at LHC.
 Indeed, new multiquark hadrons happen to be associated with at least one or two heavy quarks. 

While multiquark configurations may be heavier than the
ones made of distinct baryons, they still may constitute
 a virtual admixture to those. 
 We know from experiment that protons have ``sea" with antiquarks, in their parton composition.
 Consider alpha particle, 
 $^4He=ppnn$ state, quite deeply bound and compact according
 to nuclar physics standards. Yes, it is a ``doubly magic" one, 
 with proton and neutron lowest shell closed. But can it be mixed, in its core, with a 12-q $u^6d^6$ S-shell cluster, also a ``magic" state, in color-flavor-spin. We will return to this issue in section 
\ref{sec_9_12}.

Theory of mutiquark hadrons had rather 
controversial history. Some
ideas, which looked quite natural at the start, turned out to be rather misleading. 
The first approach  accounting for color confinement 
was represented by various ``bag models", e.g. the 
famous one from the MIT group. Hadrons were pictured as ``bubbles
of perturbative vacuum" located inside the lower ``true nonperturbative vacuum". The energy one has to pay for its creation
was represented by the ``bag constant" times volume $B\cdot V$,  accounting for the different  energy density of the two vacua.

If so, the following problem arises. As certain energy $BV$
was already used to create a bubble for a meson or baryon, 
putting more quarks into it should be energetically beneficial.  If the perturbative vacuum inside is already ``empty" of nonperturbative fields,  little penalty would come from 
adding extra quarks.  If so, why some (or all)  nuclei 
are stable rather than collapse into 
multiquark bags?  Indeed, bubbles do   have well known tendency to coalesce.

Specifically, the MIT bag model predicted the 6-q dibaryon $H=u^2d^2s^2, J^P=0^+,I=0,$ at $2150\, MeV$, so light that 
it will decay only  by a double weak decays of both $s$ quarks \cite{Jaffe:1976yi}. (There were suggestions that such particle was even observed in cosmic rays coming from a particular star, but this was shown not to be possible   \cite{Khriplovich:1985wu} because even double beta decay will happen on the way.)  This $H$ resonance was not observed.

Another naive idea which gets popular in connection with multiquark hadrons (see e.g.  \cite{JaffeWilczek}), is known as the $diquark$ model. Indeed, two $ud$ quarks can be combined either into a ``good" diquark, with color $ \underline{3}$ and spin-isospin $ S=I=0$ ` in which perturbative spin-spin
force is attractive. Nonperturbative instanton-induced forces
are attractive as well,  making then Cooper pairs  of color-superconductivity in dense quark matter \cite{Rapp:1997zu,Alford:1997zt}. 


This induced the idea  that one can construct multiquark systems out of ``good diquarks" as
elementary building blocks. If $\Delta$ mass be approximated
just by $3M$ (masses of constituent quarks), the mass of the nucleon as $3M-B_{qq}$ with one good diquark binding, then $B_{qq}\sim 300\, MeV$. Proceeding similarly,
one may similarly estimate expected masses of multiquark hadrons. 

An extreme case of this ides  was  ``quark-diquark"symmetry  model \cite{Shuryak:2003zi} with a  ``baryon-meson symmetry" and beyond . If the diquark binding can be crudely approximating as $B_{qq}\approx M$, in which 
nucleon (octet) should be approximately degenerate with mesons $$3M-B_{qq}\approx 2M$$
and hexaquarks made of three diquarks with (decuplet) baryons $$6M-3B_{qq}\approx 3M$$ 

Proceeding further with such logic all the way to 12-q state, one may predict that
states made of six ``good diquarks" have mass
$$12\cdot M- 6\cdot B_{qq}\approx 6M \approx 2M_\Delta $$
which is much lower than four nucleon masses $$4(3\cdot M-B_{qq})$$
So, such ``diquark models" would also predict
collapse of  nuclei into multiquark states. 

 (Yes, diquarks are not color neutral and there are also color confining forces between them, but, like with the bag model this generates energy growing with
quark number as power smaller than one, also leading to eventual collapse. )

Moreover, if ``good diquarks" are treated as separate elementary objects, they are scalar bosons. Therefore their   WFs should be symmetric under permutations, unlike that of quarks. So, if all of these assumptions be true,  nuclei (and $^4He$ specifically) should not exist at all. Obviously, such models are wrong, missing something very important.

A construction using ``good diquarks" as building blocks can only be a reasonable approximation if these diquarks are far from each other. Yet it is completely unclear why such configuration may be
dominant. Furthermore, 6 quarks have 15 quark pairs.
For 12 quarks there are 66 pairs. Using
most attractive channels in just 3 (or 6)  of them, and ignoring interactions of all
other quark pairs is indeed very misleading. 
Experience with atomic and nuclear shell model tells us that,
if possible, all fermions will sit at the same 1S shell states.
Yes, one has to construct  the WFs with correct Fermi symmetry (as we will do).  Its energy can be calculated only by
adding $all$ pairwise forces (to say nothing about three-body
forces and so on) between them. 

An instructive  case are tetraquarks, which gets discussed more lately,
especially due to discovery of all-charm $cc\bar c \bar c$ set of resonances at LHC. Those can be of two structures, either made of two ``good" $\underline{3}3$ or two ``bad" $6\underline{6} $ diquarks. 
Naive diquark ideology suggest that the former case should lead to lower energy. However
in the latter case the attractive color forces {\em between diquarks} are stronger (two QCD string rather than one). Including  all 6 pair-wise interactions proportional to relative color ($\vec\lambda_i \vec\lambda_j$)
one finds \cite{Badalian:2023qyi,Miesch:2023hjt} that both structures lead to the $same$ binding!

Another instructive example to be discussed below has been 
provided by \cite{Kim_2020} (KKO), for $uuuddd, I=0,S=3$ hexaquark state. 
Out of Young tableaux these authors constructed $five$ distinct mix-symmetry
WFs (we will discuss them below), and combined those
into $unique$ wf with 
total (Fermi-required) antisymmetry. Contrary to naive  expectations,
whether they do contain ``good diquarks" or not, their energies
from ($\vec\lambda \vec\lambda$) Hamiltonian turned out to be the $same$. And indeed, which quarks we decided to pair first
is a completely random choice, out of many possibilities. 

In this paper we will not use  pairing of some particular quarks into pre-selected quantum numbers, but directly construct the pertinent (anti)symmetric  WFs  based on
 representation of the permutation groups $S_n$. They will be shown to lead directly to explanations of 
the
two examples just discussed.

\section{Multiquark WFs and representations of the symmetric groups}
This paper is quite technical, so we try to make it as pedagogical as possible. While starting from 
general strategy and defining the needed steps, we also
show how they work for well known
 cases. 

{\bf The monom basis and spin-tensors:} A quark has a color index $a=1,2,3$, a spin $i=1,2$. We will consider various options for flavor, starting from $SU(2),u,d$,
then $SU(3),u,d,s$ etc. Even for a simpler former case, a single quark has
$3\cdot 2 \cdot 2=12$ states, and then for $n$ quarks a complete basis of all possible ``monoms" has $12^n$ elements
$$ |c_1...c_n f_1... f_n s_1...s_n >$$
Even for baryons, $n=3$, it is rather large,  while the number of nonzero elements are often small enough to simply list those.
But  this number  grows rapidly with $n$, so it is rather impractical to continue doing so. Standard spin-tensor notations
and usage of software such as Mathematica make it uniform and
practical for $n=4,5,6..,$.

Throughout this paper we use various spin-tensor notations, in 
which the WF components are numerated by quantum states of each
quark, but in certain predetermined order, e.g.
\be \Psi=\sum C(c_1...c_n f_1... f_n s_1...s_n)|c_1...c_n f_1... f_n s_1...s_n >\ee
with each $c_i=1,2,3$ being color indices, $s_i=1,2$ the spin indices and $f_i$ corresponding to quark flavors, $ud$ or $uds$ etc. In Mathematica one can use $Flatten$ command to 
eliminate brackets of such tensor and reduce the WF to a single one-dimensional vector (list). Going back from it to original
tensor is also relatively simple.

\subsection{Outline of Procedure}
\label{sec_procedure}

The proposed procedure for obtaining the WFs of  given multiquark hadron consistent with Fermi statistics
can be summarized with the following  steps:
\begin{enumerate}
    \item Find  {\bf possible tensor structures} that WFs should take in $each$ sector. Write any one tensor that obeys all the correct symmetries of the corresponding Young Tableaux.  For example in color sector, one can choose antisymmetric $\epsilon_{c1,c2,c3}$ 
    for baryons, with color indices $c1,c2,c3=1,2,3$
    For hexaquarks it generalizes to two of them
    $\epsilon_{c1,c2,c3}\epsilon_{c4,c5,c6}$, with all permutations of indices. If antiquarks are present, there can be also 
    be Kroneker $\delta_{cq}^{c\bar q}$ symbols included. 
    \item Write {\bf all possible  permutations} of that object's $n$ indices under generators   of $S_n$.  Express those $n!$ tensors as vectors over the basis of all $N^n$ monoms spanning $\mathbb{S}^{N^n}$.
    \item {\bf Orthogonalize} this list of vectors, using a procedure such as Gram-Schmidt.  This will result in an new orthonormal basis, typically being much shorter than the original list.  To predict exactly how much shorter, one can use the explicit formulas for this $n'$ given in equations \ref{eq_su2_dim} and \ref{eq_su3_dim}.
    \item Find the matrices that correspond to each of the 2 {\bf permutation group generators} acting on this basis $\mathbb{S}^{N^n}$ (see \ref{sec_big_permute}) 
    to obtain 2 $n'\times n'$ matrices.
    \item After steps $1-4$ for each sector (color,flavor,spin, orbital...) are done,
    take the {\bf Kronecker (tensor) product} of all sectors' permutation generator matrices, to find the effects of these permutation generators in the total Hilbert space. This yields two square matrices of dimension $n'_{total}=n'_{color}\cdot n'_{isospin}\cdot \dots$.
    \item {\bf Diagonalize}  both generators, looking for eigenvectors with with the same required symmetry (eigenvalues $1$ or $-1$) depending on the case.  (see \ref{sec_simultaneous_diag}). If found,  these eigenstates are the WFs. They can be projected back to the original monom basis, if desired.   
    \end{enumerate}

\section{Baryons} \label{sec_baryons}
In mathematics groups usually are defined via geometrical settings,
e.g. the symmetry group $S_3$ of free elements defined  as a self-maps of the equilateral triangle, $S_4$ by that of tetrahedron, and $S_6$ of 5-dimensional construction with 6 corners. 
(In fact, 
specific kinematics based on light cone description of the 
corresponding wave functions in momentum representation
lead precisely to WFs defined on exactly those geometrical objects.)

Starting with baryons, $n=3$, the well known  simplification
is that the $color$ wave function  takes  the form of Levi-Civita 
antisymmetric symbol $$\epsilon_{abc}=LeviCivita[3][[a,b,c]]\, (Mathematica)$$ . It factorizes from the rest and is antisymmetric. Therefore, the rest of the WF --
spin, flavor and orbital parts -- should  be symmetric 
under particle interchanges.

It is simple to achieve for  $\Delta$ (and other  members of $SU(3)$ decuplet) by making $each$ of them (flavor and spin) WFs $symmetric$ individually.
Say, when isospins and spins are all up, $I_z,S_z=3/2 $ the WF
of such $\Delta^{++} $
is reduced to a single flavor-spin monom $(u^\uparrow u^\uparrow u^\uparrow)$
(out of potential $2^3*2^3=64$ possible components). 

\begin{figure}[h!]
    \centering
    \includegraphics[width=5cm]{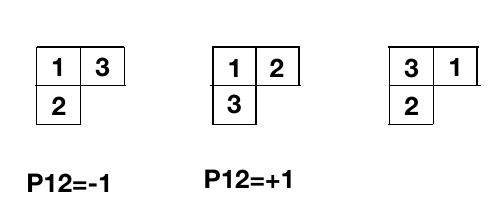}
    \caption{Three Young symmetry diagrams for  three quark spins combined  into total $S=1/2$}
    \label{fig_Young3}
\end{figure}

Yet already the nucleons (and other  members of $SU(3)$ octet) 
are  more complicated since their quantum numbers ($I,S=1/2)$ prevent construction of $separately$ symmetric WFs for spin and flavor. Three permutations of corresponding Young tableau are shown in Fig.\ref{fig_Young3}. Recall that any pair placed vertically mean antisymmetric convolution
$\sim \epsilon_{ij} S1^i S2^j, i,j=1,2 $ into zero spin pair, so the
first diagram tells us that total spin is that of quark 3, $S3$. The second does so for quarks 1 and 3, and the third to quarks 3,2. 

(Note that convolution with epsilon is necessary, and the symmetric WF 
$\sim (\uparrow \uparrow \downarrow +\uparrow  \downarrow \uparrow +\downarrow \uparrow \uparrow ) $ while have total $S_z=1/2$ correspond
to total spin $S=3/2$ and therefore to different Young tableau, with 3 horizontal squares.)

The first tableau from Fig.\ref{fig_Young3} is antisymmetric
in respect to $P_{12}$ permutation, the second is symmetric under it. The third is antisymmetric over 2-3 permutation, but 
its transformation under 1-2 are more complicated and can in general be included as some matrix. The step 1 of our procedure leads to realization
that the third term is linear combination of the first two, 
 thus one gets a reduction from $2^3=8$ monoms in spin to only 2.

Let us show steps of our suggested program, to ensure consistency with
other applications. Consider spin WF (isospin is the same). There are $2^3=8$ monoms.  Using our Mathematica command one writes and obtains back
the following: \\
\begin{widetext}
\begin{verbatim}
In:    GoodBasis[TensorProduct[LeviCivitaTensor[2], u]]
Out:   {{0, 0, 1/Sqrt[2], 0, -(1/Sqrt[2]), 0, 0, 0},
       {0, Sqrt[2/  3], -(1/Sqrt[6]), 0, -(1/Sqrt[6]), 0, 0, 0}}
\end{verbatim}
\end{widetext}
The input  $TensorProduct[LeviCivitaTensor[2],u]$ includes $all$  permutations of particles, $u=(1,0)$ is the spin-up elementary vector. The $GoodBasis$ command orthogonalizes these possible vectors,
and tells us that all permutations produce only $two$ mutually independent
and orthogonal 8-d vectors. From that point on we work only with such ``good basis" states, and spin-flavor WFs will be written as 2*2=4 dimensional, instead
of 8*8=64-dimensional. (While the corresponding simplifications does not look like much, below we will see that similar steps for other cases would reduce space of few millions
monoms to those of just about a hundred dimensions.) 

At this point we made a digression from our suggested procedures and remind  what was done historically, starting from 1970's and widely used (e.g.  in \cite{Isgur:1978xj}).
Let us for a moment consider not spin,color, and flavors, but
just coordinates/momenta of three quarks.
 For  total momentum fixed,  one needs just two  Jacobi coordinates (or momenta) traditionally called
\ba \vec\rho &=& (\vec r_1-\vec r_2)/\sqrt{2} \\
\vec\lambda &=&(\vec r_1+\vec r_2-2\vec r_3)/\sqrt{6} \nonumber
\ea
Note that they happen to be exactly the same combinations of coordinates as spins
in our ``good basis".
Note further that 
the first of those, $\rho$, is antisymmetric under $P_{12}$ permutation, while the second is symmetric under it. Thus  $two$ spin combinations 
\ba S_{\rho}={1 \over \sqrt{2}}(\uparrow \downarrow- \downarrow \uparrow) \uparrow\\
S_{\lambda}={1 \over \sqrt{6}}
(\uparrow\downarrow\uparrow+\downarrow\uparrow\uparrow-2\uparrow\uparrow\downarrow)
\ea
 were introduced. 
The same are definitions for flavor, $F_\rho,F_\lambda$, with obvious
change from spin-up to $u$, and spin-down to $d$ quark.

Complete matrices of
  permutations are
\bea
\label{eqn_perm_12}
P_{12}\times
\begin{pmatrix}
\rho\\
\lambda
\end{pmatrix}
&=&
\begin{pmatrix}
-1&0\\
0 & 1
\end{pmatrix}
\begin{pmatrix}
\rho\\
\lambda
\end{pmatrix}\nonumber\\
P_{23}\times
\begin{pmatrix}
\rho\\
\lambda
\end{pmatrix}
&=&
\begin{pmatrix}
\frac 12&\frac{\sqrt 3}2\\
\frac{\sqrt{3}}2& -\frac 12
\end{pmatrix}
\begin{pmatrix}
\rho\\
\lambda
\end{pmatrix}
\ea  
because $\rho,\lambda$ doublet transformation under $P_{12}$ is simply antisymmetric and symmetric, so the corresponding matrix is diagonal.
One can
easily construct spin-1/2 WFs symmetric (or antisymmetric) under interchange
of quarks 1 and 2 ($P_{12}$), yet they are not symmetric under other (e.g. 
$P_{23}$) permutations.  The combinations 
 $S_{\rho}F_{\rho}$ and $S_{\lambda}F_{\lambda}$ terms are both symmetric under $P_{12}$, but not under other permutations.   Looking for their superposition people guessed that $S_{\rho}F_{\rho}+S_{\lambda}F_{\lambda}$ is in fact
 symmetric under all elements of $S_3$.
 Note that this approach has unfortunate element of guessing, and even if
 positive result for a guess is obtained it is not yet clear if other
 successful solutions may exist.

In our recent paper \cite{Miesch:2023hvl} we addressed the problem 
a bit differently. 
A Kroneker product of spin and flavor combinations in basis $\rho\rho,\rho\lambda,\lambda\rho,\lambda\lambda$
is $4\times 4$ matrix. $P_{12}\times P_{12}$  is also diagonal, with two eigenvalues 1 and two -1.  Kroneker $squares$ (products of a matrix 
to itself)  
($MM=KroneckerProduct[M, M]$ in Mathematica language)
was diagonalized and its eigenstates with eiegenvalue 1 (which corresponds
to symmetric wave function) found. Then we located eigenstate $common$
to both $P_{12}$ and $P_{23}$, producing
therefore the required permutation-symmetric spin-flavor WF of the nucleon.

Now, returning to procedure advocated in this work, the only little modification is needed: instead of 
the second operator used above, $P_{23}$, one should use the second generator of the group, $P_{cyclic}$. Its Kronecker Product
to itself is the following $4\times 4$ matrix
$$
P_{cyclic}=\begin{pmatrix}
{1/4, \sqrt{3}/4, \sqrt{3}/4, 3/4} \\
{-(\sqrt{3}/4), 1/4, -(3/4), \sqrt{3}/4}\\
  {-(\sqrt{3}/4), -(3/4), 1/4, \sqrt{3}/4}\\ 
   {3/ 4, -(\sqrt{3}/4), -(\sqrt{3}/4), 1/4}
\end{pmatrix} $$
Diagonalizing it (using $Eigensystem[P_{cyclic}]$ command) one then finds that $two$ of its 4 eigenvalues
are 1, with eigenvectors being $\{1, 0, 0, 1\}, \{0, -1, 1, 0\} $
One can then observe that only $one$ of them (the first) is common
to both permutations (and in fact to all elements of the $S_3$ group). 

In \cite{Miesch:2023hvl} we generalized this approach to the WFs of the
(negative parity) $L=1$  $N^*$ resonances. Those 
were not  explicitly defined in 1970's, by Isgur and Karl
(who used instead certain limits of the WFs of strange baryons instead).
Excited baryons with $L\neq 0$ have 
 orbital
WFs depending on angles of Jacobi cooedinatess
, which also may have various permutation symmetries. Those depend on  quark coordinates linearly (quadratically,etc). As we already discussed, the Jacobi coordinates $\rho,\lambda$ have 
the same permutation symmetries and chosen spin and flavor states.   So, to get WF for $L=1$
baryons one has to find the tensor product of three copies $KronekerProduct[M,M,M]$ (then four, etc), diagonalize these $2^3\times 2^3$ matrices, look for common eigenvectors
of all permutations with the eigenvalue 1 and found them, uniquely in 
most cases.

\section{Hexaquarks and $S_6$ representations  } 
\subsection{The $uuuddd$ Hexaquark with the maximal spin $S=3$}
\label{sec_S3_hex}

Jumping now to multiquark hadrons we start with this special case. First of all, it has been seen experimentally \cite{WASA-at-COSY:2011bjg} as a resonance $d^*(2380)$ in reaction
$$ p+n \rightarrow d \pi^0 \pi^0$$ 
 Its small width $\Gamma_{d^*}\approx 70\, Mev$ is significantly below that of Delta baryon $\Gamma_\Delta\approx 115\, MeV$. This fact  was used against its interpretation as a $\Delta \Delta$ bound state. Also the $\Delta \Delta$ binding needs then to be $\approx 84\, MeV$, perhaps too large for a ``molecular" state. 

Note further, that spin and isospin (for $u,d$) are the same $SU(2)$ structure,  one can interchange them.  So, 
one  expects its mirror image with $I=3,S=0$ and the same mass.
Similar consideration will be true for other hexaquarks. 

Theory wise, as we already noted in Introduction, it is at the moment the only hexaquark state for which
 full antisymmetric WF has actually been
derived \cite{Kim:2020rwn} and will be mentioned below as the KKO WF. That has been done by traditional means,
starting from diquarks and then adding up representations of colors and flavors to six. The answer required a lot of work and is written explicitly.

Simplification in this case comes from  the total spin being at its maximal value  $S=3$, so that all quark spins
$\uparrow\uparrow\uparrow\uparrow\uparrow\uparrow$
point in the same direction. Thus  the spin
part of the WF is trivially symmetric and factorises. What remains to deal with are
the intermixed  $color$ and $flavor$ WFs. The former make representations of the $SU(3)$ group, and the latter either $SU(2),u,d$ or $SU(3),u,d,s$ flavor groups.  

\begin{figure}
    \centering
    \includegraphics[width=8cm]{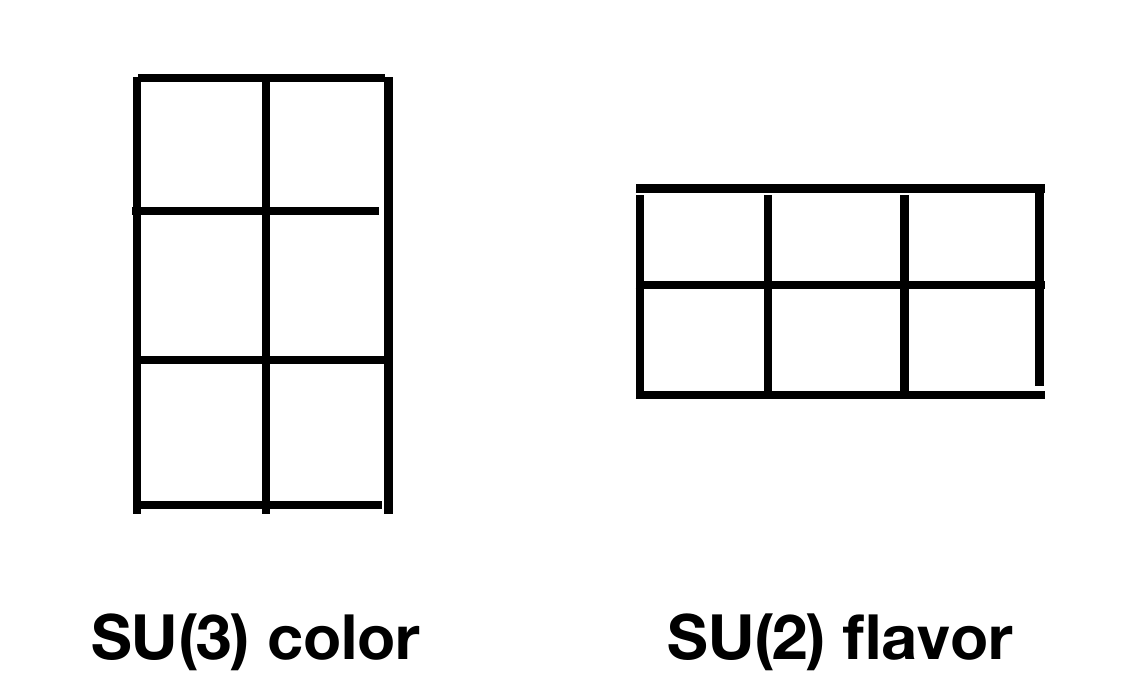}
    \caption{Color and flavor Young tableax for hexaquarks}
    \label{fig_young_6}
\end{figure}

The space of all color states have $3^6$ monoms. 
Color Young tableaux shown in Fig.\ref{fig_young_6}
should look like two complete vertical sets of squares. In our notations they correspond to all permutations of
\begin{verbatim}
 TensorProduct[LeviCivitaTensor[3], LeviCivitaTensor[3]]  
\end{verbatim}
%
For step 2, we considered every possible one of these rearrangements and wrote them in as vectors in $\mathbb{C}^{3^6}$, using Mathematica's $Flatten$ function.  There are $6!$ permutations in symmetric $S_6$ group,  so that is the number of $3^6$-dimensional vectors in our list. (By a coincidence $6!=720$  happens to be close to $3^6=729$ .)

Orthogonalization of this set of vectors was accomplished through Mathematica's $Orthogonalize$ procedure, which by default uses Gram-Schmidt method and generates a set of independent orthonormal vectors.  From dimension 720 that reduces to ``good basis" of only $5$ linearly independent combinations.  The notations below use $k=1$ is $P_{12}$,$k=2$ is $P_{cycle}$ permutation generators, defined in this basis $\{\mathbf{b}_i\}_{i=1}^5$ were then found by taking
$$
    (P^k_{color})_{ij}=\mathbf{b}_i^{T}\cdot P^k_{\mathbb{C}^{729}}\cdot \mathbf{b}_j,
$$
where $P^k_{\mathbb{C}^{729}}$ was generated using the technique described in Appendix \ref{sec_big_permute}.

For the flavor sector for $uuuddd$ quarks, the necessary tensor structure is three diquarks
\begin{widetext}
\begin{verbatim}
 TensorProduct[LeviCivitaTensor[2], LeviCivitaTensor[2],LeviCivitaTensor[2]]  
\end{verbatim}
%
Just like for color, there are $6!$ quark permutations of the flavor indices $f_i$'s, also leading to just 5 unique linearly independent combinations.  We then
compute two $P^k_{flavor}$  $5\times5$ matrices for generators of $S_6$.

The next step is to perform $KronekerProduct$ of color and flavor spaces, and
representing two
 generators of the $S_6$ group, $P_{12}$ and $P_{cycle}$. Their diagonalization allows to search for common eigenstates with total eigenvalue $-1$ (antisymmetry for Fermions).
There is indeed  one such state found. 

Most of the wavefunctions we obtained are too large to list here, but for instance the color-flavor 25-dimensional wavefunction can be written in the (``flattened") basis $\{\mathbf{b}_{color}\}\otimes\{\mathbf{b}_{flavor}\}$
\bea
    (0, 0, 0, 0, \frac{1}{\sqrt{5}}, 0, 0, 0, -\frac{1}{\sqrt{5}}, 0, 0, 0, 
    \frac{1}{\sqrt{5}}, 0, 0, 
0, \frac{1}{\sqrt{5}}, 0, 0, 0, -\frac{1}{\sqrt{5}}, 0, 0, 0, 0)
\eea

Note that there are 3 positive terms and 2 negative terms with otherwise equal contributions: we checked that they are exactly those  found by Kim, Kim, 
and Oka in \cite{Kim:2020rwn}.

\begin{table}[]
    \centering
    \begin{tabular}{|c|cccc|}
          $S$\textbackslash $S_z$ & 0 & 1 & 2 & 3   \\
         \hline
          0 & 0 &&&\\
          1 & $\{1,5\times 5\times 9\}$ & $\{1,5\times 5\times 9\}$ & & \\
          2 & 0 & 0 &  &\\
          3 & $\{1,5\times 5\times 1\}$ & $\{1,5\times 5\times 1\}$ & 
          $\{1,5\times 5\times 1\}$ & $\{1,5\times 5\times 1\}$ 
    \end{tabular}
    \caption{Hexaquarks with $uuuddd$ flavor content, with different
    values of spin $S$
    (rows) and 
    its projections $S_z$ (columns).
    When the number of states is 0, as it is for $S=0$ and 2, the row is all zeroes. When the number of antiperiodic wave functions found is one, we give  the dimension of the orthogonalized basis for tensor product of color-flavor-spin states. }
    \label{tab:udududTableSmS}
\end{table}
\end{widetext}
\subsection{Hexaquarks with arbitrary spin, flavor and  orbital momenta $L=0,1,2$} \label{sec_other_hex}

Using the power of the proposed method, we apply it for 
hexaquarks with other quantum numbers.  


For spins smaller than 3, the spin WF is no longer trivial.
For $S=1, S_z=0$, the general symmetry structure can be obtained from Young tableaux.  It is antisymmetry in 2 pairs of indices and symmetry in the other two.  The tensor structure to be permuted is therefore $[\uparrow\downarrow][\uparrow\downarrow] \{\uparrow\downarrow\}$, where 
symmetric and antisymmetric  combinations are
$$[\uparrow\downarrow]=(\uparrow\downarrow-\downarrow\uparrow)/\sqrt{2}$$
$$\{\uparrow\downarrow\}=(\uparrow\downarrow+\downarrow\uparrow)/\sqrt{2}$$
 with large number of permutations of indices. Yet, 
after $orthogonalization$ the basis of independent states  possesses only 9 linearly independent vectors. Therefore, two $S_6$ generators  $\sigma^1_{spin}$ and $\sigma^2_{spin}$  in this basis are $9\times 9$ matrices, given as examples in equation \ref{eq:9x9}.

\begin{widetext}
    \bea
        P^1_{spin}&=&\left(
        \begin{array}{ccccccccc}
         -1 & 0 & 0 & 0 & 0 & 0 & 0 & 0 & 0 \\
         0 & -1 & 0 & 0 & 0 & 0 & 0 & 0 & 0 \\
         0 & 0 & -1 & 0 & 0 & 0 & 0 & 0 & 0 \\
         0 & 0 & 0 & 1 & 0 & 0 & 0 & 0 & 0 \\
         0 & 0 & 0 & 0 & 1 & 0 & 0 & 0 & 0 \\
         0 & 0 & 0 & 0 & 0 & 1 & 0 & 0 & 0 \\
         0 & 0 & 0 & 0 & 0 & 0 & 1 & 0 & 0 \\
         0 & 0 & 0 & 0 & 0 & 0 & 0 & 1 & 0 \\
         0 & 0 & 0 & 0 & 0 & 0 & 0 & 0 & 1 \\
        \end{array}
        \right),\nonumber\\
        P^2_{spin}&=&\left(
        \begin{array}{ccccccccc}
         \frac{1}{4} & \frac{1}{4 \sqrt{3}} &
           \frac{1}{\sqrt{6}} & -\frac{\sqrt{3}}{4} &
           -\frac{1}{4} & -\frac{1}{\sqrt{2}} & 0 & 0 & 0
           \\
         -\frac{\sqrt{3}}{4} & \frac{1}{12} & \frac{1}{3
           \sqrt{2}} & -\frac{1}{4} & \frac{1}{12
           \sqrt{3}} & \frac{1}{3 \sqrt{6}} &
           -\frac{\sqrt{\frac{2}{3}}}{3} & -\frac{4}{3
           \sqrt{3}} & 0 \\
         0 & -\frac{\sqrt{2}}{3} & \frac{1}{6} & 0 &
           -\frac{\sqrt{\frac{2}{3}}}{3} & \frac{1}{6
           \sqrt{3}} & -\frac{1}{3 \sqrt{3}} & \frac{1}{6
           \sqrt{6}} & -\frac{\sqrt{\frac{5}{2}}}{2} \\
         -\frac{\sqrt{3}}{4} & -\frac{1}{4} &
           -\frac{1}{\sqrt{2}} & -\frac{1}{4} &
           -\frac{1}{4 \sqrt{3}} & -\frac{1}{\sqrt{6}} &
           0 & 0 & 0 \\
         \frac{3}{4} & -\frac{1}{4 \sqrt{3}} &
           -\frac{1}{\sqrt{6}} & -\frac{1}{4 \sqrt{3}} &
           \frac{1}{36} & \frac{1}{9 \sqrt{2}} &
           -\frac{\sqrt{2}}{9} & -\frac{4}{9} & 0 \\
         0 & \sqrt{\frac{2}{3}} & -\frac{1}{2 \sqrt{3}} &
           0 & -\frac{\sqrt{2}}{9} & \frac{1}{18} &
           -\frac{1}{9} & \frac{1}{18 \sqrt{2}} &
           -\frac{\sqrt{\frac{5}{6}}}{2} \\
         0 & 0 & 0 & \sqrt{\frac{2}{3}} &
           -\frac{\sqrt{2}}{9} & -\frac{4}{9} &
           -\frac{1}{9} & -\frac{2 \sqrt{2}}{9} & 0 \\
         0 & 0 & 0 & 0 & \frac{8}{9} & -\frac{2
           \sqrt{2}}{9} & -\frac{1}{9 \sqrt{2}} &
           \frac{1}{36} & -\frac{\sqrt{\frac{5}{3}}}{4}
           \\
         0 & 0 & 0 & 0 & 0 & 0 &
           \frac{\sqrt{\frac{3}{10}}}{2}+\frac{7}{2
           \sqrt{30}} &
           \frac{\sqrt{\frac{3}{5}}}{4}-\frac{2}{\sqrt{15
           }} & -\frac{1}{4} \\
        \end{array}
        \right)
    \label{eq:9x9}
    \eea
\end{widetext}

The next step is to define these two generators ($k=1,2$ for $P_{12}$ and $P_{cycle}$)  written as tensor product incorporating every sector of the WF, e.g. 
\be
    P^k_{total}=P^k_{color}\otimes P^k_{flavor}\otimes P^k_{spin}.
\ee

In the previous subsection -- hexaquarks with maximal spin $S=3$ -- these were $5\cdot 5\cdot 1=25$-dimensional matrices
of permutations. For other spin values and $uuuddd$ quarks, those we found to be matrices in the following minimal dimensions:
 for $S=2$ they are  in 125-d space, for $S=1$ in 225-d, and for spin 0 they are matrices in 125-dimensions again. 

While all of them are too large to be given here, we still emphasize that these
dimensions are many times  smaller than that of the full  space of monoms, $12^6$.
Important that in practice there is absolutely no problem to operate with them
inside Mathematica. In particularly, all are  generated in a second,  and 
diagonalized as  quickly.  For the details of  simultaneous diagonlization and procedure to find common antisymmetric eigenstates see Appendix \ref{sec_simultaneous_diag}.

 The particular number of solutions for each spin  depends only on the spin value $S$, and of course not on its projection $S_z$, as follows from rotational symmetry. Yet the calculation themselves are not technically identical, so 
 we did it for all values of $S_z$ to check for their mutual consistency. 
 Some of the results are shown in Table \ref{tab:udududTableSmS}.
 
 At $S=0$ and $S=2$ we found no solutions were possible with the permutation antisymmetry desired.  At $S=1$ and $S=3$ however, we found a
 $single$ antiperiodic wave function for each value of $S_z$. 

\begin{table}[]
    \centering
    \begin{tabular}{|c|cccc|}
        \hline 
        $I=0$&&&&\\
      L \textbackslash \, S   &0 & 1 & 2 & 3   \\
         \hline
         0&0& 1& 0 & 1\\
         1& 1&1 & 2 &0\\
         2&4&9&5&2 \\ \hline
    \end{tabular}
    \begin{tabular}{|c|cccc|}
        \hline 
        $I=1$&&&&\\
      L \textbackslash \, S   &0 & 1 & 2 & 3   \\
         \hline
         0&1& 0& 1 & 0\\
         1& 1&4 & 2 &1\\
         2&9&15&10&2 \\ \hline
    \end{tabular}
    \begin{tabular}{|c|cccc|}
        \hline 
        $I=2$&&&&\\
      L \textbackslash \, S   &0 & 1 & 2 & 3   \\
         \hline
         0&0& 1& 0 & 0\\
         1& 2&2 & 1 &0\\
         2&5&10&5&1 \\ \hline
    \end{tabular}
    \caption{Number of antisymmetric 6 $q$ states (per choice of $m_s$ and $m_l$) at each combination of total orbital and spin angular momentum for the light quark hexaquark. For example, last in row 1 of the $I=0$ plot is the KKO state with S=3, L=0.  At $L=0$ there are 3 possible combinations of $S$ and $I$, with the ability to swap them identically as they are both $SU(2)$: (1,0), (3,0), and (2,1).  It is worth noting that these exactly match the allowed quantum number combinations for dibaryons made from nucleons and $\Delta$'s first calculated in 1964 by Dyson and Xuong\cite{PhysRevLett.13.815}.}
    \label{tab:ududud}
\end{table}
\begin{table}[]
    \centering
    \begin{tabular}{|c|cccc|}   \hline
         L \textbackslash \, S &0 & 1 & 2 & 3   \\
         \hline
         0&0& 1& 0 & 0\\
         1& 0&2 & 1 &0\\
         2&5&7&4&1 \\ \hline
    \end{tabular}
    \caption{Number of $udsuds$ antisymmetric states (per choice of $m_s$ and $m_l$) at each combination of total orbital and spin angular momentum for $udsuds$ hexaquark.  --}
    \label{tab:udsuds}
\end{table}

Let us now change the flavor content, adding two strange quarks to $uuddss$ hexaquark. The flavor becomes $SU(3)$ and its treatment is similar to 
that of color, if total adds to zero. The resulting antisymmetric states are reported in Table \ref{tab:udsuds}.

Completing the hexaquark discussion, let us consider another simplified case, of same-flavor quarks (e.g. $cccccc$). 
The number of good states is in the Table \ref{tab:cccccc}.

Let us explain some cases without solutions first.
If both flavor and spin is flat-symmetric, then Fermi statistics requirement falls on color
WF, which cannot be fulfilled for 6 quarks.

\begin{table}[]
    \centering
    
    \begin{tabular}{|c|cccc|}  \hline 
         L \textbackslash \, S&0 & 1 & 2 & 3   \\
         \hline
         0&1& 0& 0 & 0\\
         1& 0&1 & 0 &0 \\
         2& 2& 2& 1& 0 \\  \hline
    \end{tabular}
    \caption{Number of antisymmetric states (per choice of $m_s$ and $m_l$) at each combination of total orbital and spin angular momentum for $cccccc$ hexaquark.}
    \label{tab:cccccc}
\end{table}


\section{Pentaquarks} \label{sec_penta}
\subsection{Pentaquark components of baryons}
In atomic and nuclear physics it is well known that
any state can be viewed as the lowest state in some
mean field potential, plus infinite (but convergent) sum over 
particle-hole pairs. The same is true for hadrons, in particular
baryon wave function includes the basis 3-q sector, plus 5-q sector with an extra quark-antiquark pair, etc. 

Such description is especially natural in light cone formulation, where one of the central physics issues is calculation
of the ``antiquark sea" and its observed  flavor asymmetry, 
large difference between the $\bar u$ and $\bar d$ PDfs of the proton.

Most popular description of the ``antiquark sea" is done via a combination of approaches,
such as most traditional DGLAP (gluon-based $\bar q q$ production) ,
or pion-based $\bar q q$ production.  In our own paper \cite{Shuryak:2022wtk}  quark pair production is attributed to four-fermion t'Hooft instanton-based Lagrangian. Common to all of them is 
 $incoherent$ (or probabilistic kinetic) approximation, $assuming$ that no interference between the produced and original quarks occurs, so that one can calculate $probability$ of quark pair production as if it happens in empty space.
 
 Strictly speaking, all of these mechanisms should be
 treated coherently, by adding pertinent operators to the
 Hamiltonian, connecting 5-q and 3-q sectors dynamically, and only then calculating additions to the wave function.
 Yes, there will be ``mostly 3-q" and ``mostly 5-q" states,
 after Hamiltonian gets diagonalized.

While we leave this ambitions program for future,
  in this paper we focus on CM frame and 5-q component alone.
  In this sector we set antiquark aside (e.g. think of it be heavy $c,b$) and focus on WFs of the
 four quarks, getting the wave function as required by Fermi statistics. So, the symmetry group considered in this section is $S_4$.

\subsection{$udud\overline{Q}$ pentaquarks}
  The simplest way to approach this is to find the wavefunction of the four quarks first, and then Clebsch-Gordon the relevant antiquark wavefunction on to the result.

The pentaquark color symmetry will be antisymmetric in three indices, and the last index will be controlled by the color of the antiquark.  The starting tensor with this symmetry we chose was $\epsilon_{c_1 c_2 c_3} \delta_{c_4 1}$, where without loss of generality (because color is never directly observed) we have chosen the color of the antiquark to correspond to the first color basis vector.

After writing and then orthogonalizing every flattened rearrangement of this tensor, we found 3 linearly independent elements.

For the $udud\overline{q}$ pentaquark a tensor of the flavor of the quarks can be written $\epsilon_{f_1 f_2}\epsilon_{f_3 f_4}$.  There are 2 linearly independent combinations of its rearrangements.

For spin, we must consider the possible representations of the four quarks first.  They will fall into classifications with integer total spins 0, 1, or 2.  Just as with the hexaquark, the starting tensor we choose for a given combination of total and projected spin will be the product of some symmetrized and some antisymmetrized basis spinors, i.e for $S=1,$ $S_z=1$ $[\uparrow\downarrow]\uparrow\uparrow$.  

For coordinate angular momentum, we use the same Jacobi 
coordinates, but only look at the representations of the generators of $S_4$ with them.  The first generator $(1$ $2)$ is the same, but the second one is (1 2 3 4) instead of (1 2 3 4 5).

Once we have the permutation matrices for each sector, we can take their Kronecker product and then simultaneously diagonalize them to find the allowed states for the first four quarks.  The multiplicity of states for each combination of total spin and orbital angular momentum is given in table \ref{tab:ududnoq}.
\begin{table}[]
    \centering
    \begin{tabular}{c|ccc}
      L \textbackslash \, S   &0 & 1 & 2   \\
         \hline
         0&0& 1& 0 \\
         1& 2&3 & 1\\
         2&8&12&4
    \end{tabular}
    \caption{Number of antisymmetric $udud\overline{q}$ states (per choice of $m_s$ and $m_l$) at each combination of total orbital and spin angular momentum before Clebsch-Gordoning on the last antiquark.}
    \label{tab:ududnoq}
\end{table}

After these states are found, the full system's wavefunctions can be found multiplying by the relevant multiple by the $S=\frac12$ doublet.  For instance, a single $S=1$ state of 4 quarks becomes an $S=\frac12$ state and an $S=\frac32$ state of the pentaquark, because $\mathbf{3}\otimes \mathbf{2}=\mathbf{4}\oplus \mathbf{2}$.  Because the multiplicity across $S_z$ is the same in the initial representations, it remains constant across the final representations too.  The counts of each quark's final wavefunction are shown in table \ref{tab:ududq}.

\begin{table}[]
    \centering
    \begin{tabular}{c|ccc}
         L\textbackslash \, S&$\frac12$ & $\frac32$ & $\frac52$  \\
         \hline
         0&1& 1& 0 \\
         1& 5&4 & 1 \\
         2& 20 & 16 & 4
    \end{tabular}
    \caption{Number of antisymmetric states (per choice of $m_s$ and $m_l$) at each combination of total orbital and spin angular momentum for $udud\overline{q}$ pentaquark after accounting for the antiquark.}
    \label{tab:ududq}
\end{table}
\begin{table}[]
    \centering
    \begin{tabular}{c|ccc}
         L\textbackslash \, S&$\frac12$ & $\frac32$ & $\frac52$  \\
         \hline
         0&0& 0& 0 \\
         1& 2&1 & 0 \\
         2& 9& 6 & 1
    \end{tabular}
    \caption{Number of antisymmetric states (per choice of $m_s$ and $m_l$) at each combination of total orbital and spin angular momentum for $udscq$ pentaquark.}
    \label{tab:udscq}
\end{table}



\section{Inclusion of Orbital Angular Momentum}

If one wishes to generalize the method to excited states beyond $S$ shell, it is done by an addition of a new sector: angular coordinates.  Here it helps to think of the system in terms of the modified Jacobi coordinates.   For instance in 6 dimensions the transformation to those is
\begin{widetext}
\be
\label{eq:jacobimat}
    \left(
    \begin{array}{c}
     \rho_1 \\
     \rho_2\\
     \rho_3 \\
    \rho_4 \\
    \rho_5\\
     \rho_6\\
    \end{array}
    \right)=
    \left(
    \begin{array}{cccccc}
     \frac{1}{\sqrt{2}} & 0 & 0 & 0 & 0 & 0 \\
     0 & \sqrt{\frac{2}{3}} & 0 & 0 & 0 & 0 \\
     0 & 0 & \frac{\sqrt{3}}{2} & 0 & 0 & 0 \\
     0 & 0 & 0 & \frac{2}{\sqrt{5}} & 0 & 0 \\
     0 & 0 & 0 & 0 & \sqrt{\frac{5}{6}} & 0 \\
     0 & 0 & 0 & 0 & 0 & \sqrt{\frac{6}{7}} \\
    \end{array}
    \right)
    \left(
    \begin{array}{cccccc}
     1 & -1 & 0 & 0 & 0 & 0 \\
     \frac{1}{2} & \frac{1}{2} & -1 & 0 & 0 & 0
       \\
     \frac{1}{3} & \frac{1}{3} & \frac{1}{3} &
       -1 & 0 & 0 \\
     \frac{1}{4} & \frac{1}{4} & \frac{1}{4} &
       \frac{1}{4} & -1 & 0 \\
     \frac{1}{5} & \frac{1}{5} & \frac{1}{5} &
       \frac{1}{5} & \frac{1}{5} & -1 \\
     \frac{1}{6} & \frac{1}{6} & \frac{1}{6} &
       \frac{1}{6} & \frac{1}{6} & \frac{1}{6}
       \\
    \end{array}
    \right)
    \left(
    \begin{array}{c}
     x_1 \\
     x_2\\
     x_3 \\
    x_4 \\
    x_5\\
     x_6\\
    \end{array}
    \right)
\ee
\end{widetext}

If we assume the wavefunction depends radially only on the hyperdistance sum of these coordinates $\rho^2_1+\rho^2_2+\dots+\rho_{n-1}^2$ as in \cite{}, then the angular dependence of the wavefunction can be written as a superposition of $Y_m^l(\theta_{\rho_i},\phi_{\theta_{\rho_i}})$

At $L=1$ the spherical harmonics are linear in the components of the $\rho_i$ so the transformation is simple.  If one writes the $n$-dimensional permutation $\sigma$ as a $n\times n$ matrix $S_\sigma$ in the traditional way, the effect of the two permutation generators will be 
\be
    S_{Jacobi}=U S_\sigma U^T.
\ee
Matrices for two generators of $S_6$, as
 5*5 matrices for hexaquarks, have been calculated. The tensor product to those should be
 included together other sectors during step 6, with color,flavor and spin ones.
(  The resultant projection back to monoms has to be taken  carefully, with the  understanding that $n-1$ of the overall dimensions represent Jacobi coordinates.)

At $L=2,$ instead of being linear in the cartesian coordinates of the $\rho_i$, the spherical harmonics are quadratic.  Therefore, the natural approach is to include \textit{two} permutations of the Jacobi corrdinates, $S_{Jacobi}\otimes S_{Jacobi}$, one corresponding to the first factor of the factorized quadratic and the other to the second factor.  There is still no way for a permutation to risk rotating the actual coordinates, only mixing them, so our approach of separating different values of $L_z$ is still valid.

To generalize to higher values of angular momentum, it is natural to conclude that all one must do is append more tensor products of $S_{Jacobi}$, because the $L$th order spherical harmonic is a degree $L$ polynomial in its Cartesian coordinates.  Of course, as more are matrices are added the algorithm for finding the eigenvalues increases cubicly in runtime, but this procedure is in principle applicable for any combination of angular momenta one could want.

Completing this section, we remind that experimental findings 
in which negative parity (L=1) hexaquark part of dibaryon WFs 
play some role were discussed in review \cite{Clement_2017}.

\section{Matrix Elements of basic operators and hexaquark masses}
With the color-flavor-spin wave functions available, one can
attempt to calculate the average values of pertinent operators. The obvious step one is to do that
perturbatively, for a gluon exchanges. The lowest order
gluon exchange generates potentials proportional to ``relative
color" operators made out of color generators $\langle\lambda_i^A\lambda_j^A/4\rangle$ where
$A=1..8$ and $i,j=1..n$. Relativistic corrections lead to
spin-spin, spin-orbit and tensor forces, as usual. Perturbative
one-gluon exchange require that those also are proportional to  colors, e.g. spin-spin is proportional to $$\langle\sum_{i>j}({1 \over 4}\lambda_i^A\lambda_j^A) (\vec S_i \vec S_j)\rangle$$
For $S-shell, L=0$ hadrons the spin-spin forces are the only relativistic corrections. 

The resulting masses are shown for light hexaquarks in table \ref{tab:matrixElements6q} and for all other $L=0$ states in table \ref{tab:matrixElements}.  It is interesting to note that the ordering of the six multiquark state masses in \ref{tab:matrixElements6q} (third column) does not quite match the ordering of the dibaryon molecules with the same quantum numbers (fourth and fifth column).  It is almost the same, but the KKO state uniquely breaks the pattern.  Not only is it degenerate with the (1,2) state and lighter than the (2,1) state, it is lighter than (or at least close to) the dibaron molecule state with spin 3 and isospin 0, which was predicted to have a mass of 2350 MeV.  Every other hexaquark is lighter as a molecule than a six quark ensemble, but not this particular $d'$. Perhaps this is related to the fact that this is the hexaquark state that has been experimentally observed.

\begin{figure}[h]
    \centering
    \includegraphics[width=6cm]{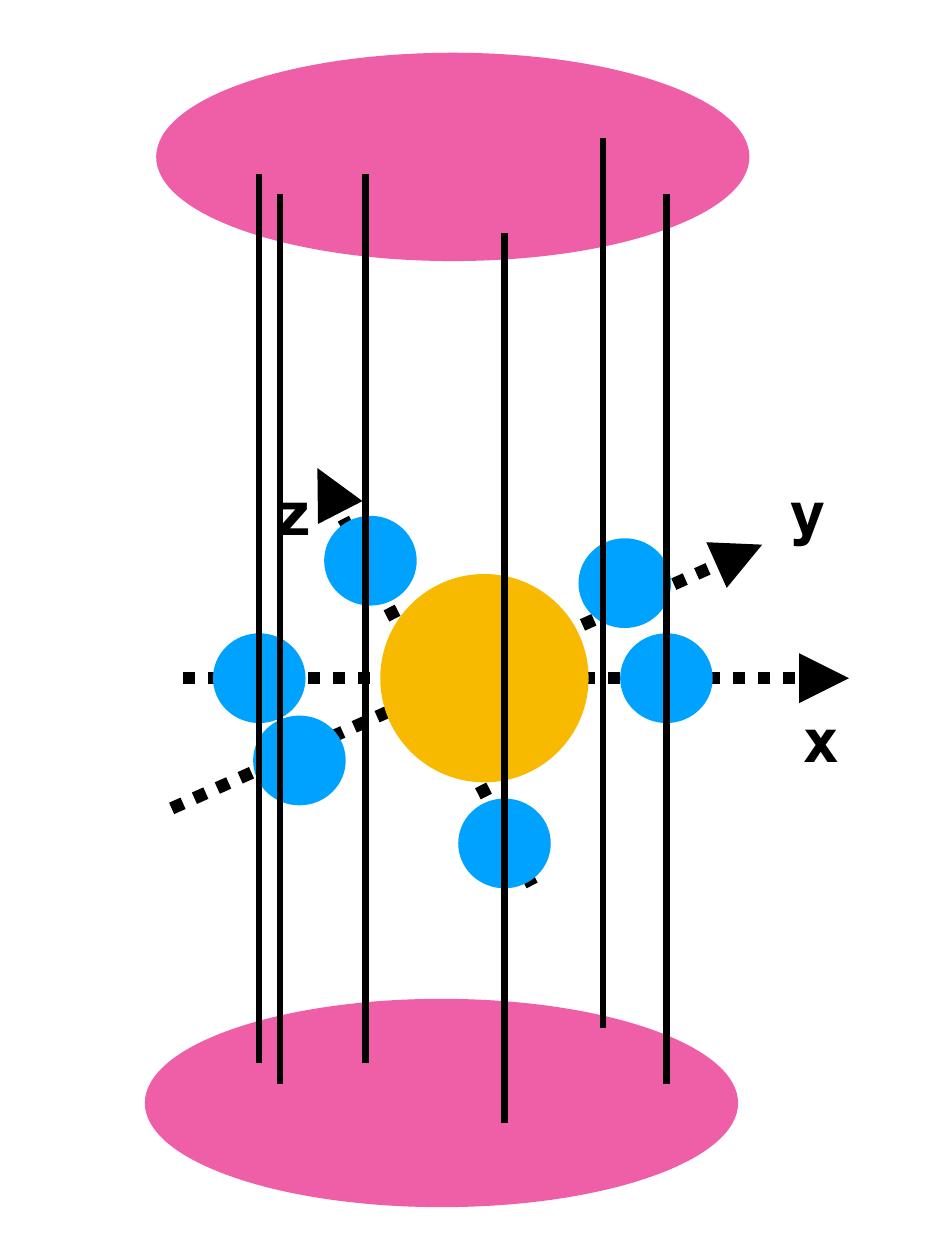}
    \caption{The setting for calculation of the effective potential for $n=6$ quarks, located at (blue) points in three dimensions. 
    The vertical direction is Euclidean time, six vertical lines are Wilson lines. Red ovals above and below indicate
    color wave functions to which Wilson lines are traced. Yellow circle indicate gauge fields of an instanton.}
    \label{fig_6W}
\end{figure}

Higher order gluon exchanges lead to operators with higher orders of Gell-Mann matrices. Those diagrams can best be obtained from an expansion of the set of Wilson lines
convoluted with color wave functions, see e.g. Fig.\ref{fig_6W} for hexaquarks.  For example, for hexaquarks one can either put two color epsilons with all $6!$  permutations or simplify it to just 5 ``good basis" color convolutions.  As far as we know, next order
gluon exchanges were not yet used in spectroscopy.

    \begin{table}[]
        \centering
        \begin{tabular}{|c|c|c|c|c|c|}
            \hline
              $(I,S)$ & $\langle\lambda\lambda\rangle$ & $\langle\lambda\lambda S S\rangle$  & 6$q$ Mass&Molec. Mass&Experiment\\
             \hline
              $(0,1)$&-16&-2/3&2098&1876&1876\\
              $(1,0)$&-16&-2&2196&1876&1878\\
              $(1,2)$&-16&-4&2342&2160&2160\\
              $(2,1)$&-16&-20/3&2536&2160&2160\\
              $(0,3)$&-16&-4&2342&2350&2380\\
              $(3,0)$&-16&-12&2926&2350&2464\\
              \hline
        \end{tabular}
        \caption{6$q$ matrix elements of color and color-spin operators for all of the $L=0$ light hexaquark states found in this paper, and the resulting masses (in MeV) obtained from the fit in \cite{Kim:2020rwn}, using $m_q=330$ MeV, $m_{s}=500$ MeV, and $m_c=1270$ MeV.  The fourth and fifth columns compare these values to the prediction from the more molecular dibaryons predicted by Dyson and Xuong in \cite{PhysRevLett.13.815}, and to the experimentally measured values of the resonances with those quantum numbers.}
        \label{tab:matrixElements6q}
    \end{table}
\begin{table}[]
        \centering
        \begin{tabular}{|c|c|c|c|}
            \hline
              State & $\langle\lambda\lambda\rangle$ & $\langle\lambda\lambda S S\rangle$  & \text{Mass (MeV)}\\
              \hline
              $udsuds$, $S=0$&-16&6&1611\\
              \hline
              $cccccc$, $S=0$&-16&-12&7819\\
              \hline
              $udud\overline{q}$, $S=\frac32$&$-\frac{20}{3}$&$7/4$&1623\\
              $udud\overline{q}$, $S=\frac12$&$-\frac{20}{3}$&$1/2$&1714\\
              \hline
              $udud\overline{c}$, $S=\frac32$&$-7.654$&$1.117$&2607\\
              $udud\overline{c}$, $S=\frac12$&$-7.654$&$1.442$&2582\\
              \hline
              $udud\overline{Q}$, $m_Q\to \infty$, $S=\frac32,\frac12$&$-8$ & $4/3$ & 1319+$m_Q$ \\
              \hline
        \end{tabular}
        \caption{Matrix elements and masses for all other $L=0$ states found in this paper, using the same fit from \cite{Kim:2020rwn}.}
        \label{tab:matrixElements}
    \end{table}
The nonperturbative confining potentials are defined via correlators of $n$ Wilson lines, 
\be \langle \Psi |\big(\delta_{c1}^{c1'}...\delta_{cn}^{cn'}- W_{c1}^{c1'} ...W_{cn}^{c1n'}\big) | \Psi'  \rangle\ee
with path-ordered exponential of color generators
$$ W=Pexp[i (\lambda^a/2) \int g A_0^a dt ]_{c1}^{c1'} $$
The setting is shown schematically
in Fig.\ref{fig_6W}

Note that while total wave functions $\Psi$ is a spin-tensor with many different indices (color, flavor, spin etc) the main perturbative $\lambda\lambda$ operator has  only color indices. So, if $\Psi$
consists of several factorizable parts 
$$ W=\sum_A C^A\psi^A_{color}\psi^A_{flavor} \psi^A_{spin} $$
one can sum over non-color indices using normalization of those wave functions and put this operator as 
``sums of squares" of the color wave functions 
$$\sum_A C_A^2\psi^A_{color}(c)\big(\delta_{c1}^{c1'}...\delta_{cn}^{cn'}- W_{c1}^{c1'} ...W_{cn}^{c1n'}\big)\psi^A_{color}(c')$$
We have calculated it explicitly for variose states. For hexaquarks, their five color wave functions
  are explicitly given in Appendix \ref{sec_color_basis_hex}, 
 and all coefficients of  $A=1..5$ are $|C^A|^2=1/5$. We also of course convoluted 
it with Wilson lines, but the expression
 is too long to be given here (can be obtained from the 
authors upon request).
This expression can be directly evaluated on the lattice,
or by any vacuum model (e.g., in the instanton model),
producing forces among quarks in all hexaquarks.


\section{Nine and twelve-quark S-shell states } \label{sec_9_12}
Simple observation tells us that $u,d$ quarks with three colors states and two spins have $2\cdot 3 \cdot 2=12$
states, and thus it should be the ``magic number" completing
the $1S,L=0$ shell.

The corresponding quantum numbers 
are those of alpha particle $^4He=ppnn$ nucleus, which is a ``double magic" one by itself, being very well bound and compact. Large literature exist discussing whether nuclei like $^{12}C,^{16}O$
do or do not include some ``alpha clusters" in their WF.
Similarly, one can ask  whether WF of alpha particles themselves should be calculated including compact 12-quark 
1S state.

Let us start from the theory of the $^4He$ WF, which of course
has a very long history originating in 1960's. Let us just mention application of
hyperdistance approximation in 9-d setting, recently borrowed
for discussion of (fully charmed) tetraquarks \cite{Badalian:2023qyi,Miesch:2023hjt}. 

Avoiding
any approximation, one can perform numerically Path Integral Monte Carlo \cite{Shuryak:1984xr}. Recently this approach has been revived
in \cite{DeMartini:2020hka} looking for ``alpha preclustering" at temperatures $T\sim 100 \, MeV$ corresponding to freezeouts in heavy ion collisions.  
Without going into detail, let us comment that the problem remains quite challenging. At one hand, six nuclear potentials
combined have deep minimum  $\approx 6 \cdot 50 \sim 300 MeV$,
yet corresponding to small fraction of configurations: but
this technical problem can still be solved by persistent 
use of Monte Carlo. And yes, the ``precluster" component
is seen in the WF, being robust enough to survive even at $T\sim 100 \, MeV$.

A non-technical issue here is the following. While nucleon scattering data, with application of the renormalization group  allowed
to fix nuclear forces at low energy $uniquely$, this is not yet
the case for repulsive ``nuclear core". (Indeed, various
sets fitted to spectra and scattering phases predict
short-distance nuclear potentials with large spread.) Six
of such potentials added naively lead to multi-GeV repulsion,
which is too large and too uncertain to believe. As many people noted before, this issue better be addressed at the quark level, as we are going to discuss.

Yet before we do so, let us make two comments. The first, as proper, is on experimental observations.
It was noted in \cite{Nikolaev} that standard theory of p-nuclei scattering based on many successful applications does not work as well for scattering on   
$^4He$. Quantities involved include total cross section, magnitude of the diffraction slope, and the location of the diffractive minimum. Inclusion of  12-q component in the WF 
with certain parameters can remedy all three, see details in \cite{Nikolaev}.
Later in \cite{Mosallem:2002ck} similar treatment was generalize to 
scattering data of the pion-$^4He$ scattering. The proposed parameters of the cluster in both papers indicate
that the 12-q ``core object" is  not so small in size.

The second comment is theoretical. Let us start, for a moment, with 
a ``good diquark model".  
We already argued in Introduction that its reasoning is wrong logically,
and  leads to phenomena never observed. There is no reason to single out some (6 in the 12q case) pairwise forces out  many
pairs (66 ): all should be included.  It can be shown why it is so on the perturbative level (see below).

At the nonpertubative level one has to explain first where strong 
diquark binding comes from. As argued in \cite{Rapp:1997zu,Alford:1997zt} the main part
of it for light $u,d,s$ quarks comes from 
the instanton-induced 't Hooft Lagrangian.
It, in turn, is the consequence of  fermion zero modes. To produce $6\cdot B_{qq}$ binding of six diquarks it would take six instantons inside the 12-q cluster. This is quite problematic to accomplished, as the instantons  in the QCD vacuum are dilute. In summary, for 12-q objects made of light $u,d,s$ quarks the
issue naturally is elevated to instanton correlations in the nonperturbative vacuum, the problem we are not ready to attack at this time. 

Let us however approach the problem assuming that quarks are
heavy enough, so that it is treatable via nonrelativistic
Schrodinger equation, with perturbative Coulomb plus (perhaps instanton-induced) potentials.  We will start with the most symmetric case, with two quark
flavors ($c,b$) possessing the same heavy mass $M$.
We wil call those 12-Q systems.
We restrict our discussion here to the basic issue addressed in this work,
namely  construction of the S-shell WFs satisfying quark Fermi statistics. 

As we have shown above, for 4,5,6 quark systems
one can construct WFs using the ``brute force" -- building explicitly the Kroneker products of the 
color-flavor-spin components of the WF and diagoninalizing 
 generators.
For 
9-Q and 12-Q cases it is not so easy to do. The total space of monoms is in the latter case $12^{12}\approx 8.9\cdot 10^{12}$ dimensional.   
The  $3^{12}$ color states is not that large, but direct reduction to  ``good basis"  needs to start with all ($9!$ or $12!$) permutations of the open-index tensors made of (3) 4 Levi-Civita
symbols, which is not practically possible to perform as such. 

Still, reducing to smaller set of permutations we were able to  find the ``good basis"
sets. For $n=9$ the color basis is 33-dimensional, and for flavor  42-dimensional. If spin is maximal $S=9/2$ and those variables are
trivialized, the Kroneker color-spin good basis space is thus $33\times 42=1386$ dimensional. Two generators of $S_9$ group were 
calculated as matrices in such dimension, with the
unfortunate conclusion that common eigenvectors with correct eigenvalue -1 does $not$ exist.

Going for non-maximal spin adds tensor product by another 42 dimensions, 
and operating in $33\times 42\times 42$ dimensions
 we were not able to do.  
Perhaps we would be able to move further 
in the subsequent publications. 

While actually computing these WFs becomes increasingly difficult as $n$ increases, by looking at tables of Kronecker coefficients such as \cite{gibsonKronecker} it is possible to predict whether these large multiquark states do or do not exist.  The results for lowest spin and isospin (1/2 or 0) up tp 24 quarks are given in the Table \ref{tab:bigmultiquarks}. In principle this technique can be used to predict the existence of any quantum number combination, up to as high a value of $n$ as mathematicians provide. 

From this table one can conclude that, for instance, some nuclei have multiquarks with quantum numbers that allow for mixing, and thus modification of their masses and cross sections, while others do not.  Two cases discussed, $n=9$ and $n=12$, both
have 1, corresponding to clusters which can mix with e.g. $^3He$ and $^4He$. That phenomenology was already discussed. The rows for $n=15-21$
give zero, and the next quark cluster existing with such quantum numbers appear only at $n=24$, or at eight nucleons.

Exploration of the existence of quark clusters  mixing with ``nuclear" states  has been 
discussed for decades, and is still  an interesting direction for future research to take. Needless to say, possible existence
of standalone ``exotic" quark states (with "hidden color") are
even more fascinating: those with heavy quarks are now appearing in experiment. We however have not 
discuss those yet in this paper, limiting flavor content to 
only light or light+strange quarks.

\begin{table}[]
    \centering
    
    \begin{tabular}{|c|c|}  \hline 
         n &States  \\
         \hline
         9&1\\
         12& 1 \\
         15&0\\
         18& 0\\  
         21&0\\
         24&1\\
         \hline
    \end{tabular}
    \caption{Computed from a table of Kronecker coefficients, the multiplicities of spin statistics-obeying $u$ and $d$ states at higher quark numbers, with $I=S=0$ for even $n$ and $I=S=1/2$ for odd (in the odd case this is per choice of $S_z$ as before).}
    \label{tab:bigmultiquarks}
\end{table}

\section{Summary}
This paper shows how one can construct  the WFs of multiquark hadrons using the representations of the $S_n$ symmetry group. The novel method we developed had started
from excited (3-q) nucleons and 4-q tetraquarks, and extended here to the 5-q pentaquarks and 6-q hexaquarks.   
Naturally, in doing so we focuses on the most symmetric cases first, extended it further whenever possible.
We also discussed physics of some 9-q and 12-q states made of $u,d$ quarks, although getting explicit Wfs of those
turn out to be numerically challenging even for our method.

This paper follows an approach of our previous work on excited baryons \cite{Miesch:2023hvl}. The main idea is that instead of building the wave function of multi-quark hadrons starting from quark pairs (mesons or diquarks), then combining their color,flavor,spin etc into corresponding tensor products, as done traditionally, one can work  directly with
the tensor products of generators of the symmetry groups $S_n$.
In many cases considered it leads to unique (or very few)
wave functions, possessing the required Fermi statistics. 

In  \cite{Miesch:2023hvl} we worked directly with spin-tensor
form of the wave functions, e.g. for $P-shell, L=1$ nucleons.
The number of components in this case were all monoms in $SU(2)$ flavor and spin, in total $2^3\times 2^3=64$, with explcit dependence of orbital part on angle of two Jacobi coordinate, $\rho,\lambda$. In this case we constructed pertinent wave functions symmetric under $S_3$ permutation,
as Fermi statistics requires.

In this work we  generalized this method to multiquark hadrons, focusing mostly on 6-q (hexaquark) and 5-q (pentaquark) states. It turned out to be possible $not$ to calculate all $n!$
permutations, but basically
construct Kronecker products of all sectors for only $two$ generators of the $S_n$ symmetry group, $P_{12}$ and 
$P_{cycle}$. 

A good representative is the spin $S=3$ hexaquark state, already discussed in literature, with explicit construction of the wave function by \cite{Kim:2020rwn}. As we have shown, our method provides
much more direct way toward it. Furthermore, we were able to
``unfreeze" spin value and derive WFs of $ud$ hexaquark with other spins, notably
$S=1$. We also show why $S=2$ and $S=0$ do not have states with required symmetry. We also constructed WFs for
pentaquarks, considering symmetry group $S_4$. The method generalizes to states with nonzero orbital momentum, deriving permutation matrices for the
set of modified Jacobi coordinates. 

The WFs obtained should be used to calculate matrix elements of various operators in the Hamiltonian. We did so only for operators $\vec\lambda_i \vec\lambda_j$ and $(\vec\lambda_i \vec\lambda_j)(\vec S_i \vec S_j)$. The resulting masses are for hexaquarks in Tables 
\ref{tab:matrixElements6q} and for all other $L=0$ states in Table \ref{tab:matrixElements}.

{\bf Acknowledgements}
This work is supported by the Office of Science, U.S. Department of Energy under Contract  No. DE-FG-88ER40388.

\appendix
\section{Symmetry groups $S_n$} \label{sec_perm_group}
Let us start reminding the strategy in our previous work on excited baryons \cite{Miesch:2023hvl} based on representations
of the 
$S_3$ group. This symmetry group
 consists of 6 elements, unity and various permutations
\bea
\label{PERM}
P_{i=1, ...,6}=I, (12), (13), (23), (123), (132)
\eea
(hope notations are self-evident). In mathematics symmetric groups usually are defined via some geometric maps,
e.g. $S_3$ as self-maps of the equilateral triangle, $S_4$ by that of tetrahedron, and so on. See section \ref{sec_Jacobi}
for more details.

The (12) and (23) permutations are then  improper (out of plane) $O(3)$ rotations 
with determinant equal to $-1$.
Other 3 permutations  are then in-plane rotations by $\frac \pi 3$ of which $P_4$ is an example. The same definition of self-maps generalizes to any $S_n$ group.

After action of any permutation the wave function
can be expressed as a matrix in the original monom basis. In our previous work  we were located states with the required permutation symmetry by
finding common eigenstates of the two of them,  $P_{12}$ and $P_{23}$. By ``common" we mean both having eigenvalues either 1 or -1, as needed. One can also calculate the commutator of those
two matrices and locate eigenstate with eigenvalue 0. After these were found, one can observe that they in fact are common to the whole group.

In this work we generalize the approach to  higher permutation groups $S_n, n=4,5,6$. Those 
have $n!$ elements, e.g. for hexaquarks we will have a group of 720 elements. It would be hard to explicitly form their matrices and check their common eigenstates. Fortunately
all elements of the permutation group of order $n$ can be generated by only two group generators, the the full $n$-cycle
$P_{cycle}$
$$\{1,\hspace{2pt}2,\hspace{2pt}3,\dots n\}\rightarrow \{2,\hspace{2pt}3,\hspace{2pt}4,\dots 1\}$$ and the original transposition $(1\hspace{2pt}2)$\cite{Bray:2007}.  Therefore, our proposed strategy would be 
to look for common eigenstate for just two matrices.
If found their common  eigenstate 
 is guaranteed to be invariant under  {\em all possible}  permutations of $S_n$.

 The dimensions for arbitrary representations of $S_n$ are given exactly by the hook length formula\cite{Frame1954TheHG}.  Because the representations of $SU(2)$ and $SU(3)$ have predictable shapes as Young tableaux, it is possible to use this formula to determine the dimensions of their corresponding permutation group representations.  For $SU(2)$ with total spin (or isospin) $s$ and $n$ quarks the dimension is given by
\bea
\label{eq_su2_dim}
    n'_{SU(2)}=\frac{2s+1}{\frac{n}{2}+s+1}\frac{n!}{(\frac{n}{2}+s)!(\frac{n}{2}-s)!}\nonumber\\
    =\frac{2s+1}{\frac{n}{2}+s+1}\begin{pmatrix}
        n \\
        \frac{n}{2}+s
    \end{pmatrix}.
\eea
For $SU(3)$ color/isospin singlets with $n$ quarks, the dimension is
\be
\label{eq_su3_dim}
    n'_{SU(3)}=\frac{2}{(\frac{n}{3}+2)(\frac{n}{3}+1)^2}\frac{n!}{(\frac{n}{3}!)^3}.
\ee

A key part of our method  involves the finding of a "good basis" on the space of monoms (step 3).  The dimension of this basis is given by these formulas, and though they may appear to have large factorial components, plotting them reveals the dimensions are all under 100 for $n<10$ and are roughly exponential for larger $n$--a significant improvement on the $n!$ basis of monoms we would otherwise have to work with.

\section{Permutations of Monoms}
\label{sec_big_permute}

The wavefunction-finding procedure requires a way to represent the 2 generators as matrices acting on the space of all monoms, so we present one possible way of finding these matricies below.  

The general superposition of the product of $n$ $\mathbf{C}^N$ basis vectors $\sum_{i_1 i_2 \dots i_n}\psi^{i_1 i_2 \dots i_n}\hat{e}_{i_1}\otimes \hat{e}_{i_2}\otimes \dots\hat{e}_{i_n}$ maps from $n$ $N$-dimensional indices to 1 $N^n$-dimensional index via the prescription $k=i_1 +i_2 N+ i_3 N^2+\dots i_n N^{n-1}$.  Essentially converting the $i$'s into the digits of an $n$-digit number in base $N$.  If one then wants to know what the effect of swapping two quarks or any other permutation $\sigma$ is on the $(\hat{e}\otimes\hat{e}\otimes \dots\hat{e})_k$ basis vector is, one just needs to write $k$ as a number in base $N$, swap the corresponding digits according to $\sigma$, and then convert the new number $k_\sigma$ back to base 10.  The total $N^n\times N^n$ dimensional permutation matrix can be written as the matrix where position $(k,k_\sigma)=1$ for all integers $k$ from 1 to $N^n$, and all other entries are 0.  For instance, the generator $P_{12}$ acting on $\mathbf{C}^{2^3}$ takes the form
\be
    \left(
    \begin{array}{cccccccc}
     1 & 0 & 0 & 0 & 0 & 0 & 0 & 0 \\
     0 & 1 & 0 & 0 & 0 & 0 & 0 & 0 \\
     0 & 0 & 0 & 0 & 1 & 0 & 0 & 0 \\
     0 & 0 & 0 & 0 & 0 & 1 & 0 & 0 \\
     0 & 0 & 1 & 0 & 0 & 0 & 0 & 0 \\
     0 & 0 & 0 & 1 & 0 & 0 & 0 & 0 \\
     0 & 0 & 0 & 0 & 0 & 0 & 1 & 0 \\
     0 & 0 & 0 & 0 & 0 & 0 & 0 & 1 \\
    \end{array}
    \right),
\ee
where, for example, position (3,5) has a 1 because in binary, $011\to 101$ under $P=(1$ $2)$.

\section{Good basis for color WFs for hexaquarks} \label{sec_color_basis_hex}
\begin{tiny}
    
Example of drastic simplification by the orthogonalization
procedure used. For hexaquarks there are $3^6$ values of color indices (monoms) and $6!$ permutations of indices in $LeviChivita[3][[c1,c2,c3]]*LeviChivita[3][[c4,c5,c6]]$.
Orthogonalization procedure leads to only $five$ ``GoodBasis"
states, which in the monom basis are the following five orthonormal vectors

\begin{verbatim} 
{{0, 0, 0, 0, 0, 0, 0, 0, 0, 0, 0, 0, 0, 0, 0, 0, 0, 0, 0, 0, 0, 0, 0,
   0, 0, 0, 0, 0, 0, 0, 0, 0, 0, 0, 0, 0, 0, 0, 0, 0, 0, 0, 0, 0, 0, 
  0, 0, 0, 0, 0, 0, 0, 0, 0, 0, 0, 0, 0, 0, 0, 0, 0, 0, 0, 0, 0, 0, 0,
   0, 0, 0, 0, 0, 0, 0, 0, 0, 0, 0, 0, 0, 0, 0, 0, 0, 0, 0, 0, 0, 0, 
  0, 0, 0, 0, 0, 0, 0, 0, 0, 0, 0, 0, 0, 0, 0, 0, 0, 0, 0, 0, 0, 0, 0,
   0, 0, 0, 0, 0, 0, 0, 0, 0, 0, 0, 0, 0, 0, 0, 0, 0, 0, 0, 0, 0, 0, 
  0, 0, 0, 0, 0, 1/6, 0, -(1/6), 0, 0, 0, -(1/6), 0, 0, 0, 1/6, 0, 0, 
  0, 1/6, 0, -(1/6), 0, 0, 0, 0, 0, 0, 0, 0, 0, 0, 0, 0, 0, 0, 0, 0, 
  0, 0, 0, 0, 0, 0, 0, 0, 0, 0, 0, 0, 0, 0, 0, 0, 0, 0, 0, 0, 
  0, -(1/6), 0, 1/6, 0, 0, 0, 1/6, 0, 0, 0, -(1/6), 0, 0, 0, -(1/6), 
  0, 1/6, 0, 0, 0, 0, 0, 0, 0, 0, 0, 0, 0, 0, 0, 0, 0, 0, 0, 0, 0, 0, 
  0, 0, 0, 0, 0, 0, 0, 0, 0, 0, 0, 0, 0, 0, 0, 0, 0, 0, 0, 0, 0, 0, 0,
   0, 0, 0, 0, 0, 0, 0, 0, 0, 0, 0, 0, 0, 0, 0, 0, 0, 0, 0, 0, 0, 0, 
  0, 0, 0, 0, 0, 0, 0, 0, 0, 0, 0, 0, 0, 0, 0, 0, 0, 0, 0, 0, 0, 0, 0,
   0, 0, 0, -(1/6), 0, 1/6, 0, 0, 0, 1/6, 0, 0, 0, -(1/6), 0, 0, 
  0, -(1/6), 0, 1/6, 0, 0, 0, 0, 0, 0, 0, 0, 0, 0, 0, 0, 0, 0, 0, 0, 
  0, 0, 0, 0, 0, 0, 0, 0, 0, 0, 0, 0, 0, 0, 0, 0, 0, 0, 0, 0, 0, 0, 0,
   0, 0, 0, 0, 0, 0, 0, 0, 0, 0, 0, 0, 0, 0, 0, 0, 0, 0, 0, 0, 0, 0, 
  0, 0, 0, 0, 0, 0, 0, 0, 0, 0, 0, 0, 0, 0, 0, 0, 0, 0, 0, 0, 0, 0, 0,
   0, 0, 0, 0, 0, 0, 0, 1/6, 0, -(1/6), 0, 0, 0, -(1/6), 0, 0, 0, 1/6,
   0, 0, 0, 1/6, 0, -(1/6), 0, 0, 0, 0, 0, 0, 0, 0, 0, 0, 0, 0, 0, 0, 
  0, 0, 0, 0, 0, 0, 0, 0, 0, 0, 0, 0, 0, 0, 0, 0, 0, 0, 0, 0, 0, 0, 0,
   0, 0, 0, 0, 0, 0, 0, 0, 0, 0, 0, 0, 0, 0, 0, 0, 0, 0, 0, 0, 0, 0, 
  0, 0, 0, 0, 0, 0, 0, 0, 0, 0, 0, 0, 0, 0, 0, 0, 0, 0, 0, 0, 0, 0, 0,
   0, 0, 0, 0, 0, 0, 0, 0, 0, 1/6, 0, -(1/6), 0, 0, 0, -(1/6), 0, 0, 
  0, 1/6, 0, 0, 0, 1/6, 0, -(1/6), 0, 0, 0, 0, 0, 0, 0, 0, 0, 0, 0, 0,
   0, 0, 0, 0, 0, 0, 0, 0, 0, 0, 0, 0, 0, 0, 0, 0, 0, 0, 0, 0, 0, 0, 
  0, 0, 0, -(1/6), 0, 1/6, 0, 0, 0, 1/6, 0, 0, 0, -(1/6), 0, 0, 
  0, -(1/6), 0, 1/6, 0, 0, 0, 0, 0, 0, 0, 0, 0, 0, 0, 0, 0, 0, 0, 0, 
  0, 0, 0, 0, 0, 0, 0, 0, 0, 0, 0, 0, 0, 0, 0, 0, 0, 0, 0, 0, 0, 0, 0,
   0, 0, 0, 0, 0, 0, 0, 0, 0, 0, 0, 0, 0, 0, 0, 0, 0, 0, 0, 0, 0, 0, 
  0, 0, 0, 0, 0, 0, 0, 0, 0, 0, 0, 0, 0, 0, 0, 0, 0, 0, 0, 0, 0, 0, 0,
   0, 0, 0, 0, 0, 0, 0, 0, 0, 0, 0, 0, 0, 0, 0, 0, 0, 0, 0, 0, 0, 0, 
  0, 0, 0, 0, 0, 0, 0, 0, 0, 0, 0, 0, 0, 0, 0, 0, 0, 0, 0, 0, 0, 0, 0,
   0, 0, 0, 0, 0, 0, 0, 0, 0, 0, 0}, 
   {0, 0, 0, 0, 0, 0, 0, 0, 0, 0, 0,
   0, 0, 0, 0, 0, 0, 0, 0, 0, 0, 0, 0, 0, 0, 0, 0, 0, 0, 0, 0, 0, 0, 
  0, 0, 0, 0, 0, 0, 0, 0, 0, 0, 0, 0, 0, 0, 0, 0, 0, 0, 0, 0, 0, 0, 0,
   0, 0, 0, 0, 0, 0, 0, 0, 0, 0, 0, 0, 0, 0, 0, 0, 0, 0, 0, 0, 0, 0, 
  0, 0, 0, 0, 0, 0, 0, 0, 0, 0, 0, 0, 0, 0, 0, 0, 0, 0, 0, 0, 0, 0, 0,
   0, 0, 0, 1/(4 Sqrt[2]), 0, -(1/(4 Sqrt[2])), 0, 0, 0, 0, 0, 0, 0, 
  0, 0, 0, 0, 0, 0, 0, 0, 0, 0, 0, 0, 0, 0, -(1/(4 Sqrt[2])), 0, 0, 0,
   1/(4 Sqrt[2]), 0, 0, 0, 0, 0, 0, 0, -(1/(12 Sqrt[2])), 0, 1/(
  12 Sqrt[2]), 0, 0, 0, 1/(12 Sqrt[2]), 0, 0, 0, -(1/(12 Sqrt[2])), 0,
   0, 0, 1/(6 Sqrt[2]), 0, -(1/(6 Sqrt[2])), 0, 0, 0, 0, 0, 0, 0, 0, 
  0, 0, 0, 0, 0, 0, 0, 0, 0, 0, 0, -(1/(4 Sqrt[2])), 0, 1/(4 Sqrt[2]),
   0, 0, 0, 0, 0, 0, 0, 0, 0, 0, 0, 0, 0, 0, 0, 1/(12 Sqrt[2]), 
  0, -(1/(12 Sqrt[2])), 0, 0, 0, 1/(6 Sqrt[2]), 0, 0, 
  0, -(1/(6 Sqrt[2])), 0, 0, 0, 1/(12 Sqrt[2]), 0, -(1/(12 Sqrt[2])), 
  0, 0, 0, 0, 0, 0, 0, 0, 0, 0, 0, 0, 0, 0, 0, -(1/(4 Sqrt[2])), 0, 
  1/(4 Sqrt[2]), 0, 0, 0, 0, 0, 0, 0, 0, 0, 0, 0, 0, 0, 0, 0, 0, 0, 0,
   0, 0, 0, 0, 0, 0, 0, 0, 0, 0, 0, 0, 0, 0, 0, 0, 0, 0, 
  0, -(1/(4 Sqrt[2])), 0, 1/(4 Sqrt[2]), 0, 0, 0, 0, 0, 0, 0, 0, 0, 0,
   0, 0, 0, 0, 0, 0, 0, 0, 0, 0, 0, 1/(4 Sqrt[2]), 0, 0, 
  0, -(1/(4 Sqrt[2])), 0, 0, 0, 0, 0, 0, 0, 1/(12 Sqrt[2]), 
  0, -(1/(12 Sqrt[2])), 0, 0, 0, -(1/(12 Sqrt[2])), 0, 0, 0, 1/(
  12 Sqrt[2]), 0, 0, 0, -(1/(6 Sqrt[2])), 0, 1/(6 Sqrt[2]), 0, 0, 0, 
  0, 0, 0, 0, 0, 0, 0, 0, 0, 0, 0, 0, 0, 0, 0, 0, 0, 0, 0, 0, 0, 0, 0,
   0, 0, 0, 0, 0, 0, 0, 0, 0, 0, 0, 0, 0, 0, 0, 0, 0, 0, 0, 0, 0, 0, 
  0, 0, 0, 0, 0, 0, 0, 0, 0, 0, 0, 0, 0, 0, 0, 0, 0, 0, 0, 0, 0, 0, 0,
   0, 0, 0, 0, 0, 0, 0, 0, 0, 0, 0, 0, 0, 0, 0, 0, 0, 0, 0, 0, 1/(
  6 Sqrt[2]), 0, -(1/(6 Sqrt[2])), 0, 0, 0, 1/(12 Sqrt[2]), 0, 0, 
  0, -(1/(12 Sqrt[2])), 0, 0, 0, -(1/(12 Sqrt[2])), 0, 1/(12 Sqrt[2]),
   0, 0, 0, 0, 0, 0, 0, -(1/(4 Sqrt[2])), 0, 0, 0, 1/(4 Sqrt[2]), 0, 
  0, 0, 0, 0, 0, 0, 0, 0, 0, 0, 0, 0, 0, 0, 0, 0, 0, 0, 0, 0, 1/(
  4 Sqrt[2]), 0, -(1/(4 Sqrt[2])), 0, 0, 0, 0, 0, 0, 0, 0, 0, 0, 0, 0,
   0, 0, 0, 0, 0, 0, 0, 0, 0, 0, 0, 0, 0, 0, 0, 0, 0, 0, 0, 0, 0, 0, 
  0, 0, 0, 1/(4 Sqrt[2]), 0, -(1/(4 Sqrt[2])), 0, 0, 0, 0, 0, 0, 0, 0,
   0, 0, 0, 0, 0, 0, 0, -(1/(12 Sqrt[2])), 0, 1/(12 Sqrt[2]), 0, 0, 
  0, -(1/(6 Sqrt[2])), 0, 0, 0, 1/(6 Sqrt[2]), 0, 0, 
  0, -(1/(12 Sqrt[2])), 0, 1/(12 Sqrt[2]), 0, 0, 0, 0, 0, 0, 0, 0, 0, 
  0, 0, 0, 0, 0, 0, 1/(4 Sqrt[2]), 0, -(1/(4 Sqrt[2])), 0, 0, 0, 0, 0,
   0, 0, 0, 0, 0, 0, 0, 0, 0, 0, 0, 0, 0, 0, -(1/(6 Sqrt[2])), 0, 1/(
  6 Sqrt[2]), 0, 0, 0, -(1/(12 Sqrt[2])), 0, 0, 0, 1/(12 Sqrt[2]), 0, 
  0, 0, 1/(12 Sqrt[2]), 0, -(1/(12 Sqrt[2])), 0, 0, 0, 0, 0, 0, 0, 1/(
  4 Sqrt[2]), 0, 0, 0, -(1/(4 Sqrt[2])), 0, 0, 0, 0, 0, 0, 0, 0, 0, 0,
   0, 0, 0, 0, 0, 0, 0, 0, 0, 0, 0, -(1/(4 Sqrt[2])), 0, 1/(
  4 Sqrt[2]), 0, 0, 0, 0, 0, 0, 0, 0, 0, 0, 0, 0, 0, 0, 0, 0, 0, 0, 0,
   0, 0, 0, 0, 0, 0, 0, 0, 0, 0, 0, 0, 0, 0, 0, 0, 0, 0, 0, 0, 0, 0, 
  0, 0, 0, 0, 0, 0, 0, 0, 0, 0, 0, 0, 0, 0, 0, 0, 0, 0, 0, 0, 0, 0, 0,
   0, 0, 0, 0, 0, 0, 0, 0, 0, 0, 0, 0, 0, 0, 0, 0, 0, 0, 0, 0, 0, 0, 
  0, 0, 0, 0, 0, 0, 0, 0, 0, 0, 0, 0, 0, 0, 0, 0, 0, 0},
  {0, 0, 0, 0, 
  0, 0, 0, 0, 0, 0, 0, 0, 0, 0, 0, 0, 0, 0, 0, 0, 0, 0, 0, 0, 0, 0, 0,
   0, 0, 0, 0, 0, 0, 0, 0, 0, 0, 0, 0, 0, 0, 0, 0, 0, 0, 0, 0, 0, 0, 
  0, 0, 0, 0, 0, 0, 0, 0, 0, 0, 0, 0, 0, 0, 0, 0, 0, 0, 0, 0, 0, 0, 0,
   0, 0, 0, 0, 0, 0, 0, 0, 0, 0, 0, 0, 0, 0, 0, 0, 0, 0, 0, 0, 0, 0, 
  0, 0, 0, 0, 1/(2 Sqrt[6]), 0, 0, 0, 0, 0, -(1/(4 Sqrt[6])), 
  0, -(1/(4 Sqrt[6])), 0, 0, 0, 0, 0, 0, 0, 0, 0, -(1/(2 Sqrt[6])), 0,
   0, 0, 0, 0, 0, 0, 0, 0, 0, 0, 1/(4 Sqrt[6]), 0, 0, 0, 1/(
  4 Sqrt[6]), 0, 0, 0, 0, 0, 0, 0, 1/(4 Sqrt[6]), 0, 1/(4 Sqrt[6]), 0,
   0, 0, -(1/(4 Sqrt[6])), 0, 0, 0, -(1/(4 Sqrt[6])), 0, 0, 0, 0, 0, 
  0, 0, 0, 0, 0, 0, 0, 0, 0, 0, 0, 0, 0, 0, 0, 0, 0, 0, 0, 
  0, -(1/(4 Sqrt[6])), 0, -(1/(4 Sqrt[6])), 0, 0, 0, 0, 0, 1/(
  2 Sqrt[6]), 0, 0, 0, 0, 0, 0, 0, 0, 0, 1/(4 Sqrt[6]), 0, 1/(
  4 Sqrt[6]), 0, 0, 0, 0, 0, 0, 0, 0, 0, 0, 0, -(1/(4 Sqrt[6])), 
  0, -(1/(4 Sqrt[6])), 0, 0, 0, 0, 0, 0, 0, 0, 0, -(1/(2 Sqrt[6])), 0,
   0, 0, 0, 0, 1/(4 Sqrt[6]), 0, 1/(4 Sqrt[6]), 0, 0, 0, 0, 0, 0, 0, 
  0, 0, 0, 0, 0, 0, 0, 0, 0, 0, 0, 0, 0, 0, 0, 0, 0, 0, 0, 0, 0, 0, 0,
   0, -(1/(2 Sqrt[6])), 0, 0, 0, 0, 0, 1/(4 Sqrt[6]), 0, 1/(
  4 Sqrt[6]), 0, 0, 0, 0, 0, 0, 0, 0, 0, 1/(2 Sqrt[6]), 0, 0, 0, 0, 0,
   0, 0, 0, 0, 0, 0, -(1/(4 Sqrt[6])), 0, 0, 0, -(1/(4 Sqrt[6])), 0, 
  0, 0, 0, 0, 0, 0, -(1/(4 Sqrt[6])), 0, -(1/(4 Sqrt[6])), 0, 0, 0, 
  1/(4 Sqrt[6]), 0, 0, 0, 1/(4 Sqrt[6]), 0, 0, 0, 0, 0, 0, 0, 0, 0, 0,
   0, 0, 0, 0, 0, 0, 0, 0, 0, 0, 0, 0, 0, 0, 0, 0, 0, 0, 0, 0, 0, 0, 
  0, 0, 0, 0, 0, 0, 0, 0, 0, 0, 0, 0, 0, 0, 0, 0, 0, 0, 0, 0, 0, 0, 0,
   0, 0, 0, 0, 0, 0, 0, 0, 0, 0, 0, 0, 0, 0, 0, 0, 0, 0, 0, 0, 0, 0, 
  0, 0, 0, 0, 0, 0, 0, 0, 0, 0, 0, 0, 0, 0, 0, 0, 0, 0, 0, 0, 0, 0, 0,
   0, 0, 0, 1/(4 Sqrt[6]), 0, 0, 0, 1/(4 Sqrt[6]), 0, 0, 
  0, -(1/(4 Sqrt[6])), 0, -(1/(4 Sqrt[6])), 0, 0, 0, 0, 0, 0, 
  0, -(1/(4 Sqrt[6])), 0, 0, 0, -(1/(4 Sqrt[6])), 0, 0, 0, 0, 0, 0, 0,
   0, 0, 0, 0, 1/(2 Sqrt[6]), 0, 0, 0, 0, 0, 0, 0, 0, 0, 1/(
  4 Sqrt[6]), 0, 1/(4 Sqrt[6]), 0, 0, 0, 0, 0, -(1/(2 Sqrt[6])), 0, 0,
   0, 0, 0, 0, 0, 0, 0, 0, 0, 0, 0, 0, 0, 0, 0, 0, 0, 0, 0, 0, 0, 0, 
  0, 0, 0, 0, 0, 0, 0, 1/(4 Sqrt[6]), 0, 1/(4 Sqrt[6]), 0, 0, 0, 0, 
  0, -(1/(2 Sqrt[6])), 0, 0, 0, 0, 0, 0, 0, 0, 0, -(1/(4 Sqrt[6])), 
  0, -(1/(4 Sqrt[6])), 0, 0, 0, 0, 0, 0, 0, 0, 0, 0, 0, 1/(4 Sqrt[6]),
   0, 1/(4 Sqrt[6]), 0, 0, 0, 0, 0, 0, 0, 0, 0, 1/(2 Sqrt[6]), 0, 0, 
  0, 0, 0, -(1/(4 Sqrt[6])), 0, -(1/(4 Sqrt[6])), 0, 0, 0, 0, 0, 0, 0,
   0, 0, 0, 0, 0, 0, 0, 0, 0, 0, 0, 0, 0, 0, 0, 0, 0, 
  0, -(1/(4 Sqrt[6])), 0, 0, 0, -(1/(4 Sqrt[6])), 0, 0, 0, 1/(
  4 Sqrt[6]), 0, 1/(4 Sqrt[6]), 0, 0, 0, 0, 0, 0, 0, 1/(4 Sqrt[6]), 0,
   0, 0, 1/(4 Sqrt[6]), 0, 0, 0, 0, 0, 0, 0, 0, 0, 0, 
  0, -(1/(2 Sqrt[6])), 0, 0, 0, 0, 0, 0, 0, 0, 0, -(1/(4 Sqrt[6])), 
  0, -(1/(4 Sqrt[6])), 0, 0, 0, 0, 0, 1/(2 Sqrt[6]), 0, 0, 0, 0, 0, 0,
   0, 0, 0, 0, 0, 0, 0, 0, 0, 0, 0, 0, 0, 0, 0, 0, 0, 0, 0, 0, 0, 0, 
  0, 0, 0, 0, 0, 0, 0, 0, 0, 0, 0, 0, 0, 0, 0, 0, 0, 0, 0, 0, 0, 0, 0,
   0, 0, 0, 0, 0, 0, 0, 0, 0, 0, 0, 0, 0, 0, 0, 0, 0, 0, 0, 0, 0, 0, 
  0, 0, 0, 0, 0, 0, 0, 0, 0, 0, 0, 0, 0, 0, 0, 0, 0, 0, 0, 0, 0, 0, 0,
   0, 0}, 
   {0, 0, 0, 0, 0, 0, 0, 0, 0, 0, 0, 0, 0, 0, 0, 0, 0, 0, 0, 0,
   0, 0, 0, 0, 0, 0, 0, 0, 0, 0, 0, 0, 0, 0, 0, 0, 0, 0, 0, 0, 0, 0, 
  0, 0, 0, 0, 0, 0, 0, 0, 1/(2 Sqrt[6]), 0, -(1/(2 Sqrt[6])), 0, 0, 0,
   0, 0, 0, 0, 0, 0, 0, 0, 0, 0, 0, 0, -(1/(2 Sqrt[6])), 0, 1/(
  2 Sqrt[6]), 0, 0, 0, 0, 0, 0, 0, 0, 0, 0, 0, 0, 0, 0, 0, 0, 0, 0, 0,
   0, 0, 0, 0, 0, 0, 0, 0, 0, 0, 0, 0, 0, 0, -(1/(4 Sqrt[6])), 0, 1/(
  4 Sqrt[6]), 0, 0, 0, 0, 0, 0, 0, 0, 0, 0, 0, 0, 0, 0, 0, 0, 0, 0, 0,
   0, 0, -(1/(4 Sqrt[6])), 0, 0, 0, 1/(4 Sqrt[6]), 0, 0, 0, 0, 0, 0, 
  0, 1/(4 Sqrt[6]), 0, -(1/(4 Sqrt[6])), 0, 0, 0, 1/(4 Sqrt[6]), 0, 0,
   0, -(1/(4 Sqrt[6])), 0, 0, 0, 0, 0, 0, 0, 0, 0, 0, 0, 0, 0, 0, 0, 
  0, 0, 0, 0, 0, 0, 0, 0, 0, 0, 1/(4 Sqrt[6]), 0, -(1/(4 Sqrt[6])), 0,
   0, 0, 0, 0, 0, 0, 0, 0, 0, 0, 0, 0, 0, 0, -(1/(4 Sqrt[6])), 0, 1/(
  4 Sqrt[6]), 0, 0, 0, 0, 0, 0, 0, 0, 0, 0, 0, 1/(4 Sqrt[6]), 
  0, -(1/(4 Sqrt[6])), 0, 0, 0, 0, 0, 0, 0, 0, 0, 0, 0, 0, 0, 0, 
  0, -(1/(4 Sqrt[6])), 0, 1/(4 Sqrt[6]), 0, 0, 0, 0, 0, 0, 0, 0, 0, 0,
   0, 0, 0, 0, 0, 0, 0, 0, 0, 0, 0, 0, 0, 0, 0, 0, 0, 0, 0, 0, 0, 0, 
  0, 0, 0, 0, 0, -(1/(4 Sqrt[6])), 0, 1/(4 Sqrt[6]), 0, 0, 0, 0, 0, 0,
   0, 0, 0, 0, 0, 0, 0, 0, 0, 0, 0, 0, 0, 0, 0, -(1/(4 Sqrt[6])), 0, 
  0, 0, 1/(4 Sqrt[6]), 0, 0, 0, 0, 0, 0, 0, 1/(4 Sqrt[6]), 
  0, -(1/(4 Sqrt[6])), 0, 0, 0, 1/(4 Sqrt[6]), 0, 0, 
  0, -(1/(4 Sqrt[6])), 0, 0, 0, 0, 0, 0, 0, 0, 0, 0, 0, 0, 0, 0, 0, 0,
   0, 0, 0, 0, 0, 0, 0, 0, 0, 0, 0, 0, 0, 0, 0, 1/(2 Sqrt[6]), 0, 0, 
  0, -(1/(2 Sqrt[6])), 0, 0, 0, 0, 0, 0, 0, 0, 0, 0, 0, 0, 0, 0, 0, 0,
   0, 0, 0, 0, 0, 0, 0, 0, 0, 0, 0, 0, 0, 0, 0, -(1/(2 Sqrt[6])), 0, 
  0, 0, 1/(2 Sqrt[6]), 0, 0, 0, 0, 0, 0, 0, 0, 0, 0, 0, 0, 0, 0, 0, 0,
   0, 0, 0, 0, 0, 0, 0, 0, 0, 0, 0, 0, 0, 0, 0, -(1/(4 Sqrt[6])), 0, 
  0, 0, 1/(4 Sqrt[6]), 0, 0, 0, -(1/(4 Sqrt[6])), 0, 1/(4 Sqrt[6]), 0,
   0, 0, 0, 0, 0, 0, 1/(4 Sqrt[6]), 0, 0, 0, -(1/(4 Sqrt[6])), 0, 0, 
  0, 0, 0, 0, 0, 0, 0, 0, 0, 0, 0, 0, 0, 0, 0, 0, 0, 0, 0, 1/(
  4 Sqrt[6]), 0, -(1/(4 Sqrt[6])), 0, 0, 0, 0, 0, 0, 0, 0, 0, 0, 0, 0,
   0, 0, 0, 0, 0, 0, 0, 0, 0, 0, 0, 0, 0, 0, 0, 0, 0, 0, 0, 0, 0, 0, 
  0, 0, 0, 1/(4 Sqrt[6]), 0, -(1/(4 Sqrt[6])), 0, 0, 0, 0, 0, 0, 0, 0,
   0, 0, 0, 0, 0, 0, 0, -(1/(4 Sqrt[6])), 0, 1/(4 Sqrt[6]), 0, 0, 0, 
  0, 0, 0, 0, 0, 0, 0, 0, 1/(4 Sqrt[6]), 0, -(1/(4 Sqrt[6])), 0, 0, 0,
   0, 0, 0, 0, 0, 0, 0, 0, 0, 0, 0, 0, -(1/(4 Sqrt[6])), 0, 1/(
  4 Sqrt[6]), 0, 0, 0, 0, 0, 0, 0, 0, 0, 0, 0, 0, 0, 0, 0, 0, 0, 0, 0,
   0, 0, 0, 0, 0, 0, -(1/(4 Sqrt[6])), 0, 0, 0, 1/(4 Sqrt[6]), 0, 0, 
  0, -(1/(4 Sqrt[6])), 0, 1/(4 Sqrt[6]), 0, 0, 0, 0, 0, 0, 0, 1/(
  4 Sqrt[6]), 0, 0, 0, -(1/(4 Sqrt[6])), 0, 0, 0, 0, 0, 0, 0, 0, 0, 0,
   0, 0, 0, 0, 0, 0, 0, 0, 0, 0, 0, 1/(4 Sqrt[6]), 
  0, -(1/(4 Sqrt[6])), 0, 0, 0, 0, 0, 0, 0, 0, 0, 0, 0, 0, 0, 0, 0, 0,
   0, 0, 0, 0, 0, 0, 0, 0, 0, 0, 0, 0, 0, 0, 0, 0, 0, 1/(2 Sqrt[6]), 
  0, -(1/(2 Sqrt[6])), 0, 0, 0, 0, 0, 0, 0, 0, 0, 0, 0, 0, 0, 0, 
  0, -(1/(2 Sqrt[6])), 0, 1/(2 Sqrt[6]), 0, 0, 0, 0, 0, 0, 0, 0, 0, 0,
   0, 0, 0, 0, 0, 0, 0, 0, 0, 0, 0, 0, 0, 0, 0, 0, 0, 0, 0, 0, 0, 0, 
  0, 0, 0, 0, 0, 0, 0, 0, 0, 0, 0, 0, 0, 0, 0, 0, 0, 0}, {0, 0, 0, 0, 
  0, 0, 0, 0, 0, 0, 0, 0, 0, 0, 0, 0, 0, 0, 0, 0, 0, 0, 0, 0, 0, 0, 0,
   0, 0, 0, 0, 0, 0, 0, 0, 0, 0, 0, 0, 0, 0, 0, 0, 0, 1/(3 Sqrt[2]), 
  0, 0, 0, 0, 0, -(1/(6 Sqrt[2])), 0, -(1/(6 Sqrt[2])), 0, 0, 0, 0, 0,
   0, 0, 0, 0, 0, 0, 0, 0, 0, 0, -(1/(6 Sqrt[2])), 
  0, -(1/(6 Sqrt[2])), 0, 0, 0, 0, 0, 1/(3 Sqrt[2]), 0, 0, 0, 0, 0, 0,
   0, 0, 0, 0, 0, 0, 0, 0, 0, 0, 0, 0, 0, 0, 0, -(1/(6 Sqrt[2])), 0, 
  0, 0, 0, 0, 1/(12 Sqrt[2]), 0, 1/(12 Sqrt[2]), 0, 0, 0, 0, 0, 0, 0, 
  0, 0, -(1/(6 Sqrt[2])), 0, 0, 0, 0, 0, 0, 0, 0, 0, 0, 0, 1/(
  12 Sqrt[2]), 0, 0, 0, 1/(12 Sqrt[2]), 0, 0, 0, 0, 0, 0, 0, 1/(
  12 Sqrt[2]), 0, 1/(12 Sqrt[2]), 0, 0, 0, 1/(12 Sqrt[2]), 0, 0, 0, 
  1/(12 Sqrt[2]), 0, 0, 0, -(1/(6 Sqrt[2])), 0, -(1/(6 Sqrt[2])), 0, 
  0, 0, 0, 0, 0, 0, 0, 0, 0, 0, 0, 0, 0, 0, 0, 0, 0, 0, 1/(
  12 Sqrt[2]), 0, 1/(12 Sqrt[2]), 0, 0, 0, 0, 0, -(1/(6 Sqrt[2])), 0, 
  0, 0, 0, 0, 0, 0, 0, 0, 1/(12 Sqrt[2]), 0, 1/(12 Sqrt[2]), 0, 0, 
  0, -(1/(6 Sqrt[2])), 0, 0, 0, -(1/(6 Sqrt[2])), 0, 0, 0, 1/(
  12 Sqrt[2]), 0, 1/(12 Sqrt[2]), 0, 0, 0, 0, 0, 0, 0, 0, 
  0, -(1/(6 Sqrt[2])), 0, 0, 0, 0, 0, 1/(12 Sqrt[2]), 0, 1/(
  12 Sqrt[2]), 0, 0, 0, 0, 0, 0, 0, 0, 0, 0, 0, 0, 0, 0, 0, 0, 0, 0, 
  0, 0, 0, 0, 0, 0, 0, 0, 0, 0, 0, 0, 0, -(1/(6 Sqrt[2])), 0, 0, 0, 0,
   0, 1/(12 Sqrt[2]), 0, 1/(12 Sqrt[2]), 0, 0, 0, 0, 0, 0, 0, 0, 
  0, -(1/(6 Sqrt[2])), 0, 0, 0, 0, 0, 0, 0, 0, 0, 0, 0, 1/(
  12 Sqrt[2]), 0, 0, 0, 1/(12 Sqrt[2]), 0, 0, 0, 0, 0, 0, 0, 1/(
  12 Sqrt[2]), 0, 1/(12 Sqrt[2]), 0, 0, 0, 1/(12 Sqrt[2]), 0, 0, 0, 
  1/(12 Sqrt[2]), 0, 0, 0, -(1/(6 Sqrt[2])), 0, -(1/(6 Sqrt[2])), 0, 
  0, 0, 0, 0, 0, 0, 0, 0, 0, 0, 0, 0, 1/(3 Sqrt[2]), 0, 0, 0, 0, 0, 0,
   0, 0, 0, 0, 0, -(1/(6 Sqrt[2])), 0, 0, 0, -(1/(6 Sqrt[2])), 0, 0, 
  0, 0, 0, 0, 0, 0, 0, 0, 0, 0, 0, 0, 0, 0, 0, 0, 0, 0, 0, 0, 0, 0, 0,
   0, 0, 0, 0, 0, 0, -(1/(6 Sqrt[2])), 0, 0, 0, -(1/(6 Sqrt[2])), 0, 
  0, 0, 0, 0, 0, 0, 0, 0, 0, 0, 1/(3 Sqrt[2]), 0, 0, 0, 0, 0, 0, 0, 0,
   0, 0, 0, 0, 0, -(1/(6 Sqrt[2])), 0, -(1/(6 Sqrt[2])), 0, 0, 0, 1/(
  12 Sqrt[2]), 0, 0, 0, 1/(12 Sqrt[2]), 0, 0, 0, 1/(12 Sqrt[2]), 0, 
  1/(12 Sqrt[2]), 0, 0, 0, 0, 0, 0, 0, 1/(12 Sqrt[2]), 0, 0, 0, 1/(
  12 Sqrt[2]), 0, 0, 0, 0, 0, 0, 0, 0, 0, 0, 0, -(1/(6 Sqrt[2])), 0, 
  0, 0, 0, 0, 0, 0, 0, 0, 1/(12 Sqrt[2]), 0, 1/(12 Sqrt[2]), 0, 0, 0, 
  0, 0, -(1/(6 Sqrt[2])), 0, 0, 0, 0, 0, 0, 0, 0, 0, 0, 0, 0, 0, 0, 0,
   0, 0, 0, 0, 0, 0, 0, 0, 0, 0, 0, 0, 0, 0, 0, 0, 1/(12 Sqrt[2]), 0, 
  1/(12 Sqrt[2]), 0, 0, 0, 0, 0, -(1/(6 Sqrt[2])), 0, 0, 0, 0, 0, 0, 
  0, 0, 0, 1/(12 Sqrt[2]), 0, 1/(12 Sqrt[2]), 0, 0, 
  0, -(1/(6 Sqrt[2])), 0, 0, 0, -(1/(6 Sqrt[2])), 0, 0, 0, 1/(
  12 Sqrt[2]), 0, 1/(12 Sqrt[2]), 0, 0, 0, 0, 0, 0, 0, 0, 
  0, -(1/(6 Sqrt[2])), 0, 0, 0, 0, 0, 1/(12 Sqrt[2]), 0, 1/(
  12 Sqrt[2]), 0, 0, 0, 0, 0, 0, 0, 0, 0, 0, 0, 0, 0, 0, 0, 0, 0, 0, 
  0, -(1/(6 Sqrt[2])), 0, -(1/(6 Sqrt[2])), 0, 0, 0, 1/(12 Sqrt[2]), 
  0, 0, 0, 1/(12 Sqrt[2]), 0, 0, 0, 1/(12 Sqrt[2]), 0, 1/(12 Sqrt[2]),
   0, 0, 0, 0, 0, 0, 0, 1/(12 Sqrt[2]), 0, 0, 0, 1/(12 Sqrt[2]), 0, 0,
   0, 0, 0, 0, 0, 0, 0, 0, 0, -(1/(6 Sqrt[2])), 0, 0, 0, 0, 0, 0, 0, 
  0, 0, 1/(12 Sqrt[2]), 0, 1/(12 Sqrt[2]), 0, 0, 0, 0, 
  0, -(1/(6 Sqrt[2])), 0, 0, 0, 0, 0, 0, 0, 0, 0, 0, 0, 0, 0, 0, 0, 0,
   0, 0, 0, 0, 0, 1/(3 Sqrt[2]), 0, 0, 0, 0, 0, -(1/(6 Sqrt[2])), 
  0, -(1/(6 Sqrt[2])), 0, 0, 0, 0, 0, 0, 0, 0, 0, 0, 0, 0, 0, 0, 
  0, -(1/(6 Sqrt[2])), 0, -(1/(6 Sqrt[2])), 0, 0, 0, 0, 0, 1/(
  3 Sqrt[2]), 0, 0, 0, 0, 0, 0, 0, 0, 0, 0, 0, 0, 0, 0, 0, 0, 0, 0, 0,
   0, 0, 0, 0, 0, 0, 0, 0, 0, 0, 0, 0, 0, 0, 0, 0, 0, 0, 0, 0, 0, 0, 
  0, 0, 0}}
 \end{verbatim}  \end{tiny}
\section{Simultaneous Diagonalization}
\label{sec_simultaneous_diag}
Our procedure requires the  diagonalization of the two $C_n$ group generators, $P_{1\hspace{2pt}2}$ and $P_{cycle}$, with then identification of their common
eigenvectors with appropriate eigenvalues, -1 or 1.
 There are many techniques to compute the simultaneous eigenvectors of two matrices, but the one we used was simply to choose two random real numbers $a$ and $b$, and then diagonalize $aP_{1\hspace{2pt}2}+bP_{cycle}$, looking
for eigenvectors with eigenvalue exactly equal to $a+b$ in the symmetric case or $-a+(-1)^{n-1} b$ in the antisymmetric case.  (The $(-1)^{n-1}$ is necessary because the second generator, the $n$-cycle, is equal to the product of $n-1$ transpositions, each of which should flip the sign of the antisymmetric state once.)  Since the odds of hitting exactly these arbitrary real numbers any way other than by the vector being an eigenvector of both generators with the correct eigenvalues are extremely low, we can conclude with good confidence that these spinors satisfy the desired relation.  And of course, it is very easy to check by hand afterwards that they are the correct eigenvectors. 

\section{Using Kronecker Coefficient Tables}
\label{sec_kroneckerTables}
To determine the existence of wavefunction with large quark numbers 
one may use  the resource: \hyperlink{https://www.jgibson.id.au/articles/characters/}{https://www.jgibson.id.au/articles/characters/}.  To use this website to find if the needed Kronecker coefficient is zero or not, one first needs to write the representations of all its sectors as Young Tableaux.  The notation used  index each tableau (which correspond to what are called Specht modules) as a list of the lengths of each row of boxes in order. For instance with $S=3$ hexaquark $n=6$, the
Young tableaux are shown in Fig.\ref{fig_young_6}.
The
color tensor form $\epsilon_{c_1 c_2c_3}\epsilon_{c_4 c_4 c_6}$ in this web resource is written as $s[2,2,2]$ or s[$2^3$]. The flavor one $\epsilon_{f_1 f_2}\epsilon_{f_3 f_4}\epsilon_{f_5 f_6}$ is denoted by s[3,3] or s[$3^2$].  

Set the value of $S_n$ to the correct value of $n$--in this case 6.  This will bring up a list of irreducible representations for the group, which will include s[$2^3$] and s[$3^2$].  To view their tensor product, simply select $both$ rows  from the table. This will bring up another table containing Kroneker decompositions of  higher powers of it, in terms of exterior, symmetric, and tensor products.  In most cases the only column needed in this table is the first one, corresponding to the decomposition of just the state. The row we are looking for is the one for the totally antisymmetric representation.  This is always the final row, s[$1^n$], because that it is the Young tableau of $n$ blocks stacked vertically (known to mathematicians as the "sign" representation).  In the case considered, there is a "1" in the first column of the last row, which tells us there does exist $one$ antisymmetric state that can be built from the tensor product of this particular color and flavor combination. This is the KKO state we also explicitly found by
our procedure.

Note that there are  "0"  in many other cases of that table, meaning  no such state with those quantum numbers exist.  Numbers larger than one are very rare but present: it would imply higher multiplicities of states.

The program handles the combination of many different representations easily, though dealing with the case where flavor and spin are identical with nontrivial color is somewhat difficult because there is no way to select one rep twice and another once.  One solution is to look at the product of the color sector with s[$1^n$], and then try to find the resultant rep in the $\otimes^2 \chi$ column of the flavor/spin rep.  This works because the Kronecker coefficients are symmetric under interchanges of the representation indices.

\begin{widetext}
\section{Mathematica Code}
This section will provide a complete set of the code necessary to find these wavefunctions, and an example in the context of the $ududud$ hexaquark.

First, it is useful to have a way to write the two permutation generators $P_{\text{(1 2)}}$ and $P_{cycle}$ as $n\times n$ matrices (where $n$ is still the number of identical quarks).
\begin{verbatim}
    Permutes[n_] := 
     PermutationMatrix[#, n] & /@ GroupGenerators[SymmetricGroup[n]]
\end{verbatim}

The transformation to modified Jacobi coordinates in $n$ we used can implemented as a matrix (as in eq \ref{eq:jacobimat}) with Jacobi, and the generators can be projected into that basis with JacobiPermute.
\begin{verbatim}
    Jacobi[n_] := 
     Inverse[DiagonalMatrix[
        Table[Sqrt[i + 1]/Sqrt[i], {i, 1, n}]]] . (Table[
         Join[Table[1/i, i], Table[0, n - i]], {i, 1, n}] + 
        DiagonalMatrix[Table[-1, n - 1], 1])
    JacobiPermute[n_, i_] := 
     SparseArray[(Jacobi[n] . Permutes[n][[i]] . Inverse[Jacobi[n]])[[
       1 ;; n - 1, 1 ;; n - 1]]]
\end{verbatim}

BigPermute returns the full monom basis permutation matrices, obtained using the technique described in Appendix \ref{sec_big_permute}.
\begin{verbatim}
    ColumnSwap[n_, dim_, i_, x_] := 
    FromDigits[Permutes[n][[i]] . IntegerDigits[x, dim, n], dim]
    BigPermute[n_, dim_, i_] := 
    SparseArray[
        Table[{x + 1, ColumnSwap[n, dim, i, x] + 1} -> 1, {x, 0, 
        dim^n - 1}]]
\end{verbatim}

Now, we have the necessary tools to create PermuteInBasis, which accomplishes steps 2 through 4 of the method described in section \ref{sec_procedure}, given any starting tensor $T$ with rank equal to the number of quarks.
\begin{verbatim}
    PermuteInBasis[T_] := Module[{n, dim, tensorList, basis},
      n = TensorRank[T];
      dim = Length[T];
      tensorList = 
       Flatten /@ (Transpose[T, PermutationList[#, n]] & /@ 
          GroupElements[SymmetricGroup[n]]);
      basis = DeleteCases[Orthogonalize[tensorList], {0 ..}];
      {SparseArray[basis . BigPermute[n, dim, 1] . Transpose[basis]], 
       SparseArray[basis . BigPermute[n, dim, 2] . Transpose[basis]]}
      ]
\end{verbatim}

If one just wants the basis of minimal linearly independent combination of permutations of tensor indices, GoodBasis can be used.
\begin{verbatim}
    GoodBasis[T_] := Module[{n, dim, tensorList, basis},
      n = TensorRank[T];
      tensorList = 
       Flatten /@ (Transpose[T, PermutationList[#, n]] & /@ 
          GroupElements[SymmetricGroup[n]]);
      DeleteCases[Orthogonalize[tensorList], {0 ..}]
      ]
\end{verbatim}

For the example of nucleon flavor, it returns the Jacobi coordinates $\rho$ and $\lambda$ as expected.
\begin{verbatim}
    GoodBasis[TensorProduct[LeviCivitaTensor[2],u]]
    {{0,0,1/Sqrt[2],0,-(1/Sqrt[2]),0,0,0},{0,Sqrt[2/3],-(1/Sqrt[6]),0,-(1/Sqrt[6]),0,0,0}}
\end{verbatim}

There is one more piece missing.  Given two matrices M[[1]] and M[[2]], SymmetricStates and AntiSymmetricStates apply the simultaneous diagonalization technique described in appendix \ref{sec_simultaneous_diag}, which is step 6 in our procedure.
\begin{verbatim}
    SymmetricStates[M_] :=
     Module[
      {a, b, test},
      a = RandomReal[];
      b = RandomReal[];
      test = a M[[1]] + b M[[2]];
      Chop[Select[Transpose[Eigensystem[test]], 
         Round[#[[1]], 10^-3] == Round[a + b, 10^-3] &][[;; , 2]]]
      ]
    AntiSymmetricStates[M_, n_] := Module[
      {a, b, test},
      a = RandomReal[];
      b = RandomReal[];
      test = a M[[1]] + b M[[2]];
      Chop[Select[Transpose[Eigensystem[test]], 
         Round[#[[1]], 10^-3] == 
           Round[-a + (-1)^(n - 1) b, 10^-3] &][[;; , 2]]]
      ]
\end{verbatim}

With PermuteInBasis, AntiSymmetricStates, and JacobiPermute we now have all that is needed to find the wavefunctions for any particle.  Consider the $ududud$ hexaquark.  For step 1, we need to write the tensors with the correct index symmetries.  For spin, Young Tableaux give us the following:
\begin{verbatim}
    u = {1, 0}; d = {0, 1};
    hexspin0 = {TensorProduct[LeviCivitaTensor[2], LeviCivitaTensor[2], 
        LeviCivitaTensor[2]]};
    hexspin1 = {TensorProduct[LeviCivitaTensor[2], LeviCivitaTensor[2], 
        Symmetrize[TensorProduct[u, d]]], 
       TensorProduct[LeviCivitaTensor[2], LeviCivitaTensor[2], u, u]};
    hexspin2 = {TensorProduct[LeviCivitaTensor[2], 
        Symmetrize[TensorProduct[u, u, d, d]]],
       TensorProduct[LeviCivitaTensor[2], 
        Symmetrize[TensorProduct[u, u, u, d]]],
       TensorProduct[LeviCivitaTensor[2], 
        Symmetrize[TensorProduct[u, u, u, u]]]};
    hexspin3 = {Symmetrize[TensorProduct[u, u, u, d, d, d]],
       Symmetrize[TensorProduct[u, u, u, u, d, d]],
       Symmetrize[TensorProduct[u, u, u, u, u, d]],
       Symmetrize[TensorProduct[u, u, u, u, u, u]]};
\end{verbatim}

Color is more straightforward, always being just two Levi-Cevitas as discussed previously, and orbital angular momentum is a number of JacobiPermute's equal to $L$.  Once we have all of these tensors, we can use PermuteInBasis to make tables of the permutation generators in each basis.  The color, spin, and orbital angular momentum projections are the same for every hexaquark.
\begin{verbatim}
    colorPermute6 =
      PermuteInBasis[
       TensorProduct[LeviCivitaTensor[3], LeviCivitaTensor[3]]];
    spinPermute6 = {PermuteInBasis /@ hexspin0, 
      PermuteInBasis /@ hexspin1, PermuteInBasis /@ hexspin2, 
      PermuteInBasis /@ hexspin3}
    coordPermute6 = {{{{1}}, {{1}}}, {JacobiPermute[6, 1], 
       JacobiPermute[6, 2]},
      {KroneckerProduct[JacobiPermute[6, 1], JacobiPermute[6, 1]], 
       KroneckerProduct[JacobiPermute[6, 2], JacobiPermute[6, 2]]}}
\end{verbatim}

For $ududud$, flavor is an $SU(2)$ singlet.
\begin{verbatim}
    flavorPermute6ududud =
     PermuteInBasis[
      TensorProduct[LeviCivitaTensor[2], LeviCivitaTensor[2], 
       LeviCivitaTensor[2]]]
\end{verbatim}

The generators on the full space can now be constructed (step 5).  They may look different for different values of $S,$ $S_z,$ and $L$, so it is worth making a large table over these angular momenta and getting what we need from it later.
\begin{verbatim}
    fullPermute6ududud = 
      Table[{KroneckerProduct[colorPermute6[[1]], 
           flavorPermute6ududud[[1]], #[[1]], coordPermute6[[l + 1, 1]]], 
          KroneckerProduct[colorPermute6[[2]], 
           flavorPermute6ududud[[2]], #[[2]], 
           coordPermute6[[l + 1, 2]]]} & /@ spinPermute6[[sp + 1]], {sp, 
        0, 3}, {l, 0, 2}];
\end{verbatim}

We are now able to apply the simultaneous diagonalization technique to the two generators.  UdududStates gives a list of every possible antisymmetric $ududud$ state for a given angular momentum combination.  
\begin{verbatim}
    UdududStates[sp_, sz_, l_] := 
     AntiSymmetricStates[fullPermute6ududud[[sp + 1, l + 1, sz + 1]], 6]
\end{verbatim}

For $S=3$, $S_z=3$, $L=0$, there is only the KKO state.
\begin{verbatim}
    UdududStates[3, 3, 0]
    {{0, 0, 0, 0, -0.447214, 0, 0, 0, 0.447214, 0, 0, 0, -0.447214, 0, 0, 
    0, -0.447214, 0, 0, 0, 0.447214, 0, 0, 0, 0}}
\end{verbatim}

To obtain tables like Table \ref{tab:udududTableSmS}, one can run the following block.
\begin{verbatim}
    TableForm[
     Table[Table[Dimensions[UdududStates[sp, sz, 0]], {sz, 0, sp}], {sp, 0, 
       3}], TableHeadings -> {Range[0, 3], Range[0, 3]}]
\end{verbatim}

\end{widetext}

\bibliography{main}

\begin{thebibliography}{23}%
\makeatletter
\providecommand \@ifxundefined [1]{%
 \@ifx{#1\undefined}
}%
\providecommand \@ifnum [1]{%
 \ifnum #1\expandafter \@firstoftwo
 \else \expandafter \@secondoftwo
 \fi
}%
\providecommand \@ifx [1]{%
 \ifx #1\expandafter \@firstoftwo
 \else \expandafter \@secondoftwo
 \fi
}%
\providecommand \natexlab [1]{#1}%
\providecommand \enquote  [1]{``#1''}%
\providecommand \bibnamefont  [1]{#1}%
\providecommand \bibfnamefont [1]{#1}%
\providecommand \citenamefont [1]{#1}%
\providecommand \href@noop [0]{\@secondoftwo}%
\providecommand \href [0]{\begingroup \@sanitize@url \@href}%
\providecommand \@href[1]{\@@startlink{#1}\@@href}%
\providecommand \@@href[1]{\endgroup#1\@@endlink}%
\providecommand \@sanitize@url [0]{\catcode `\\12\catcode `\$12\catcode
  `\&12\catcode `\#12\catcode `\^12\catcode `\_12\catcode `\%12\relax}%
\providecommand \@@startlink[1]{}%
\providecommand \@@endlink[0]{}%
\providecommand \url  [0]{\begingroup\@sanitize@url \@url }%
\providecommand \@url [1]{\endgroup\@href {#1}{\urlprefix }}%
\providecommand \urlprefix  [0]{URL }%
\providecommand \Eprint [0]{\href }%
\providecommand \doibase [0]{http://dx.doi.org/}%
\providecommand \selectlanguage [0]{\@gobble}%
\providecommand \bibinfo  [0]{\@secondoftwo}%
\providecommand \bibfield  [0]{\@secondoftwo}%
\providecommand \translation [1]{[#1]}%
\providecommand \BibitemOpen [0]{}%
\providecommand \bibitemStop [0]{}%
\providecommand \bibitemNoStop [0]{.\EOS\space}%
\providecommand \EOS [0]{\spacefactor3000\relax}%
\providecommand \BibitemShut  [1]{\csname bibitem#1\endcsname}%
\let\auto@bib@innerbib\@empty
\bibitem [{\citenamefont {Miesch}\ \emph {et~al.}(2023)\citenamefont {Miesch},
  \citenamefont {Shuryak},\ and\ \citenamefont {Zahed}}]{Miesch:2023hvl}%
  \BibitemOpen
  \bibfield  {author} {\bibinfo {author} {\bibfnamefont {N.}~\bibnamefont
  {Miesch}}, \bibinfo {author} {\bibfnamefont {E.}~\bibnamefont {Shuryak}}, \
  and\ \bibinfo {author} {\bibfnamefont {I.}~\bibnamefont {Zahed}},\ }\href
  {\doibase 10.1103/PhysRevD.108.094033} {\bibfield  {journal} {\bibinfo
  {journal} {Phys. Rev. D}\ }\textbf {\bibinfo {volume} {108}},\ \bibinfo
  {pages} {094033} (\bibinfo {year} {2023})},\ \Eprint
  {http://arxiv.org/abs/2308.14694} {arXiv:2308.14694 [hep-ph]} \BibitemShut
  {NoStop}%
\bibitem [{\citenamefont {Gibson}(2021)}]{gibsonKronecker}%
  \BibitemOpen
  \bibfield  {author} {\bibinfo {author} {\bibfnamefont {J.}~\bibnamefont
  {Gibson}},\ }\href {https://www.jgibson.id.au/articles/characters/} {\enquote
  {\bibinfo {title} {Characters of the symmetric group},}\ } (\bibinfo {year}
  {2021})\BibitemShut {NoStop}%
\bibitem [{\citenamefont {Clement}(2017)}]{Clement_2017}%
  \BibitemOpen
  \bibfield  {author} {\bibinfo {author} {\bibfnamefont {H.}~\bibnamefont
  {Clement}},\ }\href {\doibase 10.1016/j.ppnp.2016.12.004} {\bibfield
  {journal} {\bibinfo  {journal} {Progress in Particle and Nuclear Physics}\
  }\textbf {\bibinfo {volume} {93}},\ \bibinfo {pages} {195–242} (\bibinfo
  {year} {2017})}\BibitemShut {NoStop}%
\bibitem [{\citenamefont {Bashkanov}\ \emph {et~al.}(2024)\citenamefont
  {Bashkanov}, \citenamefont {Watts}, \citenamefont {Clash}, \citenamefont
  {Mocanu},\ and\ \citenamefont {Nicol}}]{Bashkanov:2023yca}%
  \BibitemOpen
  \bibfield  {author} {\bibinfo {author} {\bibfnamefont {M.}~\bibnamefont
  {Bashkanov}}, \bibinfo {author} {\bibfnamefont {D.~P.}\ \bibnamefont
  {Watts}}, \bibinfo {author} {\bibfnamefont {G.}~\bibnamefont {Clash}},
  \bibinfo {author} {\bibfnamefont {M.}~\bibnamefont {Mocanu}}, \ and\ \bibinfo
  {author} {\bibfnamefont {M.}~\bibnamefont {Nicol}},\ }\href {\doibase
  10.1088/1361-6471/ad27e6} {\bibfield  {journal} {\bibinfo  {journal} {J.
  Phys. G}\ }\textbf {\bibinfo {volume} {51}},\ \bibinfo {pages} {045106}
  (\bibinfo {year} {2024})},\ \Eprint {http://arxiv.org/abs/2308.07066}
  {arXiv:2308.07066 [hep-ph]} \BibitemShut {NoStop}%
\bibitem [{\citenamefont {Isgur}\ and\ \citenamefont
  {Karl}(1978)}]{Isgur:1978xj}%
  \BibitemOpen
  \bibfield  {author} {\bibinfo {author} {\bibfnamefont {N.}~\bibnamefont
  {Isgur}}\ and\ \bibinfo {author} {\bibfnamefont {G.}~\bibnamefont {Karl}},\
  }\href {\doibase 10.1103/PhysRevD.18.4187} {\bibfield  {journal} {\bibinfo
  {journal} {Phys. Rev. D}\ }\textbf {\bibinfo {volume} {18}},\ \bibinfo
  {pages} {4187} (\bibinfo {year} {1978})}\BibitemShut {NoStop}%
\bibitem [{\citenamefont {Jaffe}(1977)}]{Jaffe:1976yi}%
  \BibitemOpen
  \bibfield  {author} {\bibinfo {author} {\bibfnamefont {R.~L.}\ \bibnamefont
  {Jaffe}},\ }\href {\doibase 10.1103/PhysRevLett.38.195} {\bibfield  {journal}
  {\bibinfo  {journal} {Phys. Rev. Lett.}\ }\textbf {\bibinfo {volume} {38}},\
  \bibinfo {pages} {195} (\bibinfo {year} {1977})},\ \bibinfo {note} {[Erratum:
  Phys.Rev.Lett. 38, 617 (1977)]}\BibitemShut {NoStop}%
\bibitem [{\citenamefont {Khriplovich}\ and\ \citenamefont
  {Shuryak}(1986)}]{Khriplovich:1985wu}%
  \BibitemOpen
  \bibfield  {author} {\bibinfo {author} {\bibfnamefont {I.~B.}\ \bibnamefont
  {Khriplovich}}\ and\ \bibinfo {author} {\bibfnamefont {E.~V.}\ \bibnamefont
  {Shuryak}},\ }\href@noop {} {\bibfield  {journal} {\bibinfo  {journal} {Sov.
  J. Nucl. Phys.}\ }\textbf {\bibinfo {volume} {43}},\ \bibinfo {pages} {858}
  (\bibinfo {year} {1986})}\BibitemShut {NoStop}%
\bibitem [{\citenamefont {Rapp}\ \emph {et~al.}(1998)\citenamefont {Rapp},
  \citenamefont {Sch\"afer}, \citenamefont {Shuryak},\ and\ \citenamefont
  {Velkovsky}}]{Rapp:1997zu}%
  \BibitemOpen
  \bibfield  {author} {\bibinfo {author} {\bibfnamefont {R.}~\bibnamefont
  {Rapp}}, \bibinfo {author} {\bibfnamefont {T.}~\bibnamefont {Sch\"afer}},
  \bibinfo {author} {\bibfnamefont {E.~V.}\ \bibnamefont {Shuryak}}, \ and\
  \bibinfo {author} {\bibfnamefont {M.}~\bibnamefont {Velkovsky}},\ }\href
  {\doibase 10.1103/PhysRevLett.81.53} {\bibfield  {journal} {\bibinfo
  {journal} {Phys. Rev. Lett.}\ }\textbf {\bibinfo {volume} {81}},\ \bibinfo
  {pages} {53} (\bibinfo {year} {1998})},\ \Eprint
  {http://arxiv.org/abs/hep-ph/9711396} {arXiv:hep-ph/9711396} \BibitemShut
  {NoStop}%
\bibitem [{\citenamefont {Alford}\ \emph {et~al.}(1998)\citenamefont {Alford},
  \citenamefont {Rajagopal},\ and\ \citenamefont {Wilczek}}]{Alford:1997zt}%
  \BibitemOpen
  \bibfield  {author} {\bibinfo {author} {\bibfnamefont {M.~G.}\ \bibnamefont
  {Alford}}, \bibinfo {author} {\bibfnamefont {K.}~\bibnamefont {Rajagopal}}, \
  and\ \bibinfo {author} {\bibfnamefont {F.}~\bibnamefont {Wilczek}},\ }\href
  {\doibase 10.1016/S0370-2693(98)00051-3} {\bibfield  {journal} {\bibinfo
  {journal} {Phys. Lett. B}\ }\textbf {\bibinfo {volume} {422}},\ \bibinfo
  {pages} {247} (\bibinfo {year} {1998})},\ \Eprint
  {http://arxiv.org/abs/hep-ph/9711395} {arXiv:hep-ph/9711395} \BibitemShut
  {NoStop}%
\bibitem [{\citenamefont {Shuryak}\ and\ \citenamefont
  {Zahed}(2004)}]{Shuryak:2003zi}%
  \BibitemOpen
  \bibfield  {author} {\bibinfo {author} {\bibfnamefont {E.}~\bibnamefont
  {Shuryak}}\ and\ \bibinfo {author} {\bibfnamefont {I.}~\bibnamefont
  {Zahed}},\ }\href {\doibase 10.1016/j.physletb.2004.03.019} {\bibfield
  {journal} {\bibinfo  {journal} {Phys. Lett. B}\ }\textbf {\bibinfo {volume}
  {589}},\ \bibinfo {pages} {21} (\bibinfo {year} {2004})},\ \Eprint
  {http://arxiv.org/abs/hep-ph/0310270} {arXiv:hep-ph/0310270} \BibitemShut
  {NoStop}%
\bibitem [{\citenamefont {Badalian}\ and\ \citenamefont
  {Simonov}(2023)}]{Badalian:2023qyi}%
  \BibitemOpen
  \bibfield  {author} {\bibinfo {author} {\bibfnamefont {A.~M.}\ \bibnamefont
  {Badalian}}\ and\ \bibinfo {author} {\bibfnamefont {Y.~A.}\ \bibnamefont
  {Simonov}},\ }\href {\doibase 10.1140/epjc/s10052-023-11590-z} {\bibfield
  {journal} {\bibinfo  {journal} {Eur. Phys. J. C}\ }\textbf {\bibinfo {volume}
  {83}},\ \bibinfo {pages} {410} (\bibinfo {year} {2023})},\ \Eprint
  {http://arxiv.org/abs/2301.13597} {arXiv:2301.13597 [hep-ph]} \BibitemShut
  {NoStop}%
\bibitem [{\citenamefont {Miesch}\ \emph {et~al.}(2024)\citenamefont {Miesch},
  \citenamefont {Shuryak},\ and\ \citenamefont {Zahed}}]{Miesch:2023hjt}%
  \BibitemOpen
  \bibfield  {author} {\bibinfo {author} {\bibfnamefont {N.}~\bibnamefont
  {Miesch}}, \bibinfo {author} {\bibfnamefont {E.}~\bibnamefont {Shuryak}}, \
  and\ \bibinfo {author} {\bibfnamefont {I.}~\bibnamefont {Zahed}},\ }\href
  {\doibase 10.1103/PhysRevD.109.014022} {\bibfield  {journal} {\bibinfo
  {journal} {Phys. Rev. D}\ }\textbf {\bibinfo {volume} {109}},\ \bibinfo
  {pages} {014022} (\bibinfo {year} {2024})},\ \Eprint
  {http://arxiv.org/abs/2308.05638} {arXiv:2308.05638 [hep-ph]} \BibitemShut
  {NoStop}%
\bibitem [{\citenamefont {Kim}\ \emph {et~al.}(2020{\natexlab{a}})\citenamefont
  {Kim}, \citenamefont {Kim},\ and\ \citenamefont {Oka}}]{Kim_2020}%
  \BibitemOpen
  \bibfield  {author} {\bibinfo {author} {\bibfnamefont {H.}~\bibnamefont
  {Kim}}, \bibinfo {author} {\bibfnamefont {K.}~\bibnamefont {Kim}}, \ and\
  \bibinfo {author} {\bibfnamefont {M.}~\bibnamefont {Oka}},\ }\href {\doibase
  10.1103/physrevd.102.074023} {\bibfield  {journal} {\bibinfo  {journal}
  {Physical Review D}\ }\textbf {\bibinfo {volume} {102}} (\bibinfo {year}
  {2020}{\natexlab{a}}),\ 10.1103/physrevd.102.074023}\BibitemShut {NoStop}%
\bibitem [{\citenamefont {Adlarson}\ \emph {et~al.}(2011)\citenamefont
  {Adlarson} \emph {et~al.}}]{WASA-at-COSY:2011bjg}%
  \BibitemOpen
  \bibfield  {author} {\bibinfo {author} {\bibfnamefont {P.}~\bibnamefont
  {Adlarson}} \emph {et~al.} (\bibinfo {collaboration} {WASA-at-COSY}),\ }\href
  {\doibase 10.1103/PhysRevLett.106.242302} {\bibfield  {journal} {\bibinfo
  {journal} {Phys. Rev. Lett.}\ }\textbf {\bibinfo {volume} {106}},\ \bibinfo
  {pages} {242302} (\bibinfo {year} {2011})},\ \Eprint
  {http://arxiv.org/abs/1104.0123} {arXiv:1104.0123 [nucl-ex]} \BibitemShut
  {NoStop}%
\bibitem [{\citenamefont {Kim}\ \emph {et~al.}(2020{\natexlab{b}})\citenamefont
  {Kim}, \citenamefont {Kim},\ and\ \citenamefont {Oka}}]{Kim:2020rwn}%
  \BibitemOpen
  \bibfield  {author} {\bibinfo {author} {\bibfnamefont {H.}~\bibnamefont
  {Kim}}, \bibinfo {author} {\bibfnamefont {K.~S.}\ \bibnamefont {Kim}}, \ and\
  \bibinfo {author} {\bibfnamefont {M.}~\bibnamefont {Oka}},\ }\href {\doibase
  10.1103/PhysRevD.102.074023} {\bibfield  {journal} {\bibinfo  {journal}
  {Phys. Rev. D}\ }\textbf {\bibinfo {volume} {102}},\ \bibinfo {pages}
  {074023} (\bibinfo {year} {2020}{\natexlab{b}})},\ \Eprint
  {http://arxiv.org/abs/2009.11983} {arXiv:2009.11983 [hep-ph]} \BibitemShut
  {NoStop}%
\bibitem [{\citenamefont {Dyson}\ and\ \citenamefont
  {Xuong}(1964)}]{PhysRevLett.13.815}%
  \BibitemOpen
  \bibfield  {author} {\bibinfo {author} {\bibfnamefont {F.~J.}\ \bibnamefont
  {Dyson}}\ and\ \bibinfo {author} {\bibfnamefont {N.-H.}\ \bibnamefont
  {Xuong}},\ }\href {\doibase 10.1103/PhysRevLett.13.815} {\bibfield  {journal}
  {\bibinfo  {journal} {Phys. Rev. Lett.}\ }\textbf {\bibinfo {volume} {13}},\
  \bibinfo {pages} {815} (\bibinfo {year} {1964})}\BibitemShut {NoStop}%
\bibitem [{\citenamefont {Shuryak}\ and\ \citenamefont
  {Zahed}(2023)}]{Shuryak:2022wtk}%
  \BibitemOpen
  \bibfield  {author} {\bibinfo {author} {\bibfnamefont {E.}~\bibnamefont
  {Shuryak}}\ and\ \bibinfo {author} {\bibfnamefont {I.}~\bibnamefont
  {Zahed}},\ }\href {\doibase 10.1103/PhysRevD.107.034027} {\bibfield
  {journal} {\bibinfo  {journal} {Phys. Rev. D}\ }\textbf {\bibinfo {volume}
  {107}},\ \bibinfo {pages} {034027} (\bibinfo {year} {2023})},\ \Eprint
  {http://arxiv.org/abs/2208.04428} {arXiv:2208.04428 [hep-ph]} \BibitemShut
  {NoStop}%
\bibitem [{\citenamefont {Shuryak}\ and\ \citenamefont
  {Zhirov}(1984)}]{Shuryak:1984xr}%
  \BibitemOpen
  \bibfield  {author} {\bibinfo {author} {\bibfnamefont {E.~V.}\ \bibnamefont
  {Shuryak}}\ and\ \bibinfo {author} {\bibfnamefont {O.~V.}\ \bibnamefont
  {Zhirov}},\ }\href {\doibase 10.1016/0550-3213(84)90401-2} {\bibfield
  {journal} {\bibinfo  {journal} {Nucl. Phys. B}\ }\textbf {\bibinfo {volume}
  {242}},\ \bibinfo {pages} {393} (\bibinfo {year} {1984})}\BibitemShut
  {NoStop}%
\bibitem [{\citenamefont {DeMartini}\ and\ \citenamefont
  {Shuryak}(2021)}]{DeMartini:2020hka}%
  \BibitemOpen
  \bibfield  {author} {\bibinfo {author} {\bibfnamefont {D.}~\bibnamefont
  {DeMartini}}\ and\ \bibinfo {author} {\bibfnamefont {E.}~\bibnamefont
  {Shuryak}},\ }\href {\doibase 10.1016/j.nuclphysa.2021.122336} {\bibfield
  {journal} {\bibinfo  {journal} {Nucl. Phys. A}\ }\textbf {\bibinfo {volume}
  {1016}},\ \bibinfo {pages} {122336} (\bibinfo {year} {2021})},\ \Eprint
  {http://arxiv.org/abs/2007.04863} {arXiv:2007.04863 [nucl-th]} \BibitemShut
  {NoStop}%
\bibitem [{\citenamefont {Dakno}\ and\ \citenamefont
  {Nikolaev}(1985)}]{Nikolaev}%
  \BibitemOpen
  \bibfield  {author} {\bibinfo {author} {\bibfnamefont {L.}~\bibnamefont
  {Dakno}}\ and\ \bibinfo {author} {\bibfnamefont {N.}~\bibnamefont
  {Nikolaev}},\ }\href@noop {} {\bibfield  {journal} {\bibinfo  {journal}
  {Nucl. Phys.}\ }\textbf {\bibinfo {volume} {A436}},\ \bibinfo {pages} {653}
  (\bibinfo {year} {1985})}\BibitemShut {NoStop}%
\bibitem [{\citenamefont {Mosallem}\ and\ \citenamefont
  {Uzhinskii}(2002)}]{Mosallem:2002ck}%
  \BibitemOpen
  \bibfield  {author} {\bibinfo {author} {\bibfnamefont {A.~M.}\ \bibnamefont
  {Mosallem}}\ and\ \bibinfo {author} {\bibfnamefont {V.~V.}\ \bibnamefont
  {Uzhinskii}},\ }\href@noop {} {\  (\bibinfo {year} {2002})},\ \Eprint
  {http://arxiv.org/abs/nucl-th/0208043} {arXiv:nucl-th/0208043} \BibitemShut
  {NoStop}%
\bibitem [{\citenamefont {Bray}\ \emph {et~al.}(2007)\citenamefont {Bray},
  \citenamefont {Conder}, \citenamefont {Leedham-Green},\ and\ \citenamefont
  {Brien}}]{Bray:2007}%
  \BibitemOpen
  \bibfield  {author} {\bibinfo {author} {\bibfnamefont {J.}~\bibnamefont
  {Bray}}, \bibinfo {author} {\bibfnamefont {M.}~\bibnamefont {Conder}},
  \bibinfo {author} {\bibfnamefont {C.}~\bibnamefont {Leedham-Green}}, \ and\
  \bibinfo {author} {\bibfnamefont {E.}~\bibnamefont {Brien}},\ }\href
  {\doibase 10.1090/S0002-9947-2011-05231-1} {\bibfield  {journal} {\bibinfo
  {journal} {Transactions of the American Mathematical Society}\ }\textbf
  {\bibinfo {volume} {363}} (\bibinfo {year} {2007}),\
  10.1090/S0002-9947-2011-05231-1}\BibitemShut {NoStop}%
\bibitem [{\citenamefont {Frame}\ \emph {et~al.}(1954)\citenamefont {Frame},
  \citenamefont {de~B.~Robinson},\ and\ \citenamefont
  {Thrall}}]{Frame1954TheHG}%
  \BibitemOpen
  \bibfield  {author} {\bibinfo {author} {\bibfnamefont {J.~S.}\ \bibnamefont
  {Frame}}, \bibinfo {author} {\bibfnamefont {G.}~\bibnamefont
  {de~B.~Robinson}}, \ and\ \bibinfo {author} {\bibfnamefont {R.~M.}\
  \bibnamefont {Thrall}},\ }\href
  {https://api.semanticscholar.org/CorpusID:124024352} {\bibfield  {journal}
  {\bibinfo  {journal} {Canadian Journal of Mathematics}\ }\textbf {\bibinfo
  {volume} {6}},\ \bibinfo {pages} {316 } (\bibinfo {year} {1954})}\BibitemShut
  {NoStop}%
\end{thebibliography}%
\end{document}